\documentclass[review]{elsarticle}
\makeatletter
\def\ps@pprintTitle{%
	\let\@oddhead\@empty
	\let\@evenhead\@empty
	\def\@oddfoot{}%
	\let\@evenfoot\@oddfoot}
\makeatother

\usepackage{lineno,hyperref}
\usepackage{amsfonts}
\usepackage{epstopdf}
\usepackage[export]{adjustbox}
\usepackage{float}
\usepackage{graphicx}
\usepackage{subfigure}
\usepackage{color}
\usepackage{bm}
\usepackage{textcomp} 
\usepackage{lscape}
\usepackage{array}
\usepackage{booktabs}
\usepackage{amsmath}
\usepackage{wrapfig,booktabs}
\let\oldAA\AA
\renewcommand{\AA}{\text{\normalfont\oldAA}}
\hypersetup{
	colorlinks=true,
	urlcolor= blue,
	citecolor=blue,
	linkcolor= blue,
	bookmarksopen=false,
}
\modulolinenumbers[5]

\begin{document}
	
	\begin{frontmatter}
		
		\title{Topological viewpoint of two-dimensional group III-V and IV-IV compounds in the presence of electric field and spin-orbit coupling by density functional theory and tight-binding model}
		
		\author{A. Baradaran}
		\ead{a.baradaran@ut.ac.ir}
		\author{M. Ghaffarian}
		\address{Department of Physics, University of Qom, Qom, Iran}
		
		\begin{abstract}
			
			Using the tight-binding model and density functional theory, the topological invariant of the two-dimensional (2D) group III-V and IV-IV compounds are studied in the absence and the presence of an external perpendicular electric field and spin-orbit coupling. It will be recognized that a critical value of these parameters changes the topological invariant of 2D graphene-like compounds. The significant effects of an external electric field and spin-orbit coupling are considered to the two-center overlap integrals of the Slater-Koster model involved in band structures, changing band-gap, and tuning the topological phase transition between ordinary and quantum spin Hall regime. These declare the good consistency between two theories: tight-binding and density functional. So, this study reveals topological phase transition in these materials. Our finding paves a way to extend an effective Hamiltonian, and may instantly clear some computation aspects of the study in the field of spintronic based on the first-principles methods.

		\end{abstract}
		
		
	\end{frontmatter}
	
	\section{\label{sec:Intro}Introduction}
	
	Topological insulators (TIs) have been considered in much researches in the last decade \cite{kane_2005_QSHE,Fu_2006_Z2,Fu_2007_Z2,hasan_2010_TI}. This innovative material is a nontrivial insulator phase that is insulating in bulk but holds stable metallic edge states, which are protected by time-reversal symmetry (TRS) and spin-orbit interaction (SOI) \cite{kane_2005_Z2QSHE,fu_2007_TIIS}. The two-dimensional (2D) TIs, which are known as quantum spin Hall (QSH) insulators, are suitable for the novel device because of robust edge states against backscattering caused by impurities \cite{wada_2011_Bi,medhi_IOP_24.2012_ESTIs,baradaran_2020}.
	
	Since the discovery of graphene \cite{graphene_2004}, graphene-like materials have achieved attention due to their familiar 2D honeycomb structure. Some of these materials have stable flat structures exactly like graphene, such as single-layer of boron pnictides \cite{Zhuang_2012_boronPnictide}, and others are stable in buckled geometry, such as silicene \cite{gian_2007_silicon-based_nanos,Trivedi_2014_silicene_Germanene}, germanene \cite{Trivedi_2014_silicene_Germanene} and stanene \cite{Modarresi_2015_stanene}. In imitation of the discovery of the TI phase in graphene as 2D Kane-Mele material induced SOI \cite{kane_2005_Z2QSHE}, the buckled monolayers of the group III-V and IV-IV such as monolayer of Si, Ge, and Sn \cite{ezawa2015}; GaAs film by tensile strain \cite{Zhao_2015_GaAsStrain}; Bi and Sb quasi-two dimensions \cite{Fukui_2007_LCN,Sawahata_2018_Bi}, and arsenene \cite{kamal_2015_arsenene} allow us to observe the QSH. Moreover, SOI in 2D materials has been known in two ways, intrinsic; and Rashba spin-orbit that can be controlled by a perpendicular electric field which is generally smaller than the intrinsic SOI \cite{rashba_1984,Manchon_2015_rashba}. The effect of these parameters on the QSH phase needs more exploring by tuning SOI or applying an EF. Density functional theory  and tight-binding (TB) model \cite{Slater_1930_STO,Slater_1954_TB} provide us suitable tools to study these effects. In particular, 2D monolayer compounds have been of precise interest to us, whose topological behavior can be investigated with the simplest TB model \cite{Chadi_1975_TB,Vogl_1983_TB} as the nearest neighbor atoms, with SOI \cite{peterson_2000} or in the presence of an EF \cite{Ast_2012_efield}. These researches inspiring for us to investigate a more comprehensive study of 2D monolayers of the group III-V and IV-IV.
	
	The binary monolayer structures of compounds of the groups III-V and IV-IV, in addition to the elements of these groups, can be formed a buckled honeycomb lattice and stabilize by cutting in the direction of (111) (Fig.~\ref{fig:1}). The atomic structure, phonon modes, mechanical properties, and electronic structure for some of these materials and their heterostructures have been studied by \ifmmode \mbox{\c{S}}\else \c{S}\fi{}ahin et al. \cite{Sahin_2009_IIIV}. Then, others showed that the electric field produces a tunable band-gap in Dirac-type electronic spectrum, and investigated in silicene \cite{Drummond_2012_silicene}. Also, the electronic structure of silicene was simulated by the TB method with basis $sp^3d^5s^*$ \cite{Gert_2015_siliceneTB}, and the effective Hamiltonian in the vicinity of Dirac point was constructed. Even, Liu et al. derived the low-energy effective Hamiltonian involving SOI
	for silicene and 2D germanium and tin \cite{Liu_2011_EFHTB}. They showed that Si and Ge monolayers, despite their LB geometrical structure, have electronic band structures that are similar to graphene. Also, in the buckled structures of stanene and silicene with hybridization $sp^3$ type \cite{Broek_2014_tin}, the hexagonal Brillouin zone (BZ) attributes a massless Dirac fermion character to the charge carriers. Other attempts were made to account for the effect of spin splitting in buckled monolayers of group-IV \cite{farzaneh_2020_14}, and the effect of an electric field of GaAs monolayer \cite{wu_2015_EFGaAs} from first-principles calculations. They also sought a robust large gap 2D topological insulators in hydrogenated III-V buckled systems \cite{Crisostomo_2015_2DTI}. However, in all these works, the monolayer of the groups III-V and IV-IV is not considered together from a topological point of view that was the starting point for our work.
	
	In this research, inspired by recent developments and exploring 2D monolayer graphene-like materials using DFT calculation, the band structures of group III-V and IV-IV compounds are obtained, and the QSH effect is studied by calculation $\mathbb{Z}_2$ from a topological viewpoint \cite{Fukui_2007_LCN} by OpenMX package \cite{OpenMX1,OpenMX2,OpenMX3,OpenMX4,Ozaki_2003_LCPAO,Ozaki_2004_LCPAO}. Since the band-gap size is effective on the topological behavior of these materials, we show the topological properties of them with the band-gap. It is well known the size of the band-gap in TIs presents an urgent challenge that could be impressible by applying external EF, tuning SOI, and strain-induced \cite{ezawa2015,kamal_2015_arsenene,Drummond_2012_silicene,wu_2015_EFGaAs}. So, we investigate the effect of perpendicular EF and SOI on the topological phase transition of some compounds. For well understanding of the effect of external EF and SOI on electronic structures of these compounds, the Slater-Koster parameters of them in TBA are extracted with fitting DFT results using fitting methods and packages \cite{nakhaii_2018,kim_2018_TBFIT}.  Therefore, we would try to explain the TB model well effective and in detail. The present work, which considers a total of $15$ binary compounds in a 2D honeycomb structure and reveals for some of these materials, such as monolayer of BSb and SiGe, a phase transition has occurred. Fitting of DFT and TB model outputs for band structures would be carried out to find the Slater-Koster overlap integrals in the TB model, even in the presence of a perpendicular EF. The origin of the band-gap and its magnitude, which plays a key role in changing the topological phase of the materials, has been carefully studied too. The work done in this paper on these materials deals with the topic of the topological insulators and integrally examines all the 2D monolayers of the group IV-IV and group III-V, while previous researches have either not addressed the topic of the topological phase or have only examined some present compounds. These efforts nourish an outline for future research of an effective Hamiltonian and can be quite promising for nanoelectronics. Therefore, this article provides a context for the spintronic applications of 2D topological graphene-like materials \cite{Han_2016_2Dspintronic} by tunning SOI or EF.

	\section{\label{sec:Met}Methods}
	
	We have performed first-principles calculations within fully relativistic DFT at the Perdew–Burke–Ernzerhof (PBE) level~\cite{Hammer_1999_PBE} for relaxation, total energy, and electronic structures via the OpenMX package~\cite{OpenMX3,OpenMX4}. All calculations were done by using the generalized gradient approximation (GGA) as the exchange-correlation functional. Norm-conserving pseudopotentials~\cite{Hamann_1979_NCPS} and the linear combination of multiple pseudo-atomic orbitals were implemented for wave function expansion~\cite{Ozaki_2003_LCPAO, Ozaki_2004_LCPAO}. The cutoff radius for the various elements used in this article, from the lightest element, B to the heaviest, Bi, is between $2.1~ a.u.$ and $2.8~a.u.$. We have set a standard choice for pseudo-atomic orbitals~\cite{PAO2019}, and 350~Ry for cutoff energy, and \textbf{k}-space sampling points of $21\times21\times1$ for the reciprocal lattice vectors. The calculations have incorporated the SOI by a $\mathit{j}$-dependent pseudopotential composed relativistic (fully relativistic pseudopotential)~\cite{Theurich_2001_jPseudo} based on the OpenMX database 2019 (PBE19)~\cite{PAO2019}. This system is considered a single-layer slab model with a vacuum of more than five times the lattice constant, which ignores the interaction between the layers.
	
	Attempts to fit the conduction bands of material with the nearest-neighbor $sp^3$ tigh-binding model have generally succeeded.  A quantum technical software package to construct a TB model has been used named Tight Binding Studio~\cite{nakhaii_2018,kim_2018_TBFIT} starting from the simplified linear combination of atomic orbitals method in combination with first-principles calculations. Also, TB parameter fitting package (TBFIT)~\cite{kim_2018_TBFIT} for the Slater-Koster method has been applied increasing the reliability and validity of the results.
	
	We have used the methods for computing the $\mathbb{Z}_2$ invariant by utilizing the lattice Chern number (LCN)~\cite{Fukui_2007_LCN} in the OpenMX package because the parity method cannot be applied to systems with EF~\cite{Fu_2007_Z2}, or with two different atoms in a unit cell, so the $\mathbb{Z}_2$ invariant have been computed by the methods applicable without inversion symmetry~\cite{Fu_2006_Z2}. $\mathbb{Z}_2=1$ (mod 2) corresponds to the topological insulator, and $\mathbb{Z}_2=0$ (mod 2) corresponds to the trivial insulator.
	
	\section{\label{sec:AtomS}Atomic structures}
	
	Graphene forms a planar (PL) honeycomb structure since it exhibits purely $sp^2$ hybridization. On the other hand, some elemental honeycomb systems that have been found so far are not planar and form a buckled structure. Instantly, silicene, germanene, stanene, and some compounds of group-IV elements form a buckled structure, also for group III-V materials. These 2D honeycomb structures can be formed by cutting the bulk crystal along (111) Miler index. These observations make it important to study if there is a stable honeycomb geometry or not. The results of optimized geometric structures for the materials are summarized in Table.\ref{tab:1}, as Geo.  and Azimuth degree ($\theta$).
	
	\begin{figure}[!h] 
		\centering
		\begin{tabular}[b]{c}
			\begin{minipage}[c]{0.4\textwidth}
				\includegraphics[width=1\textwidth]{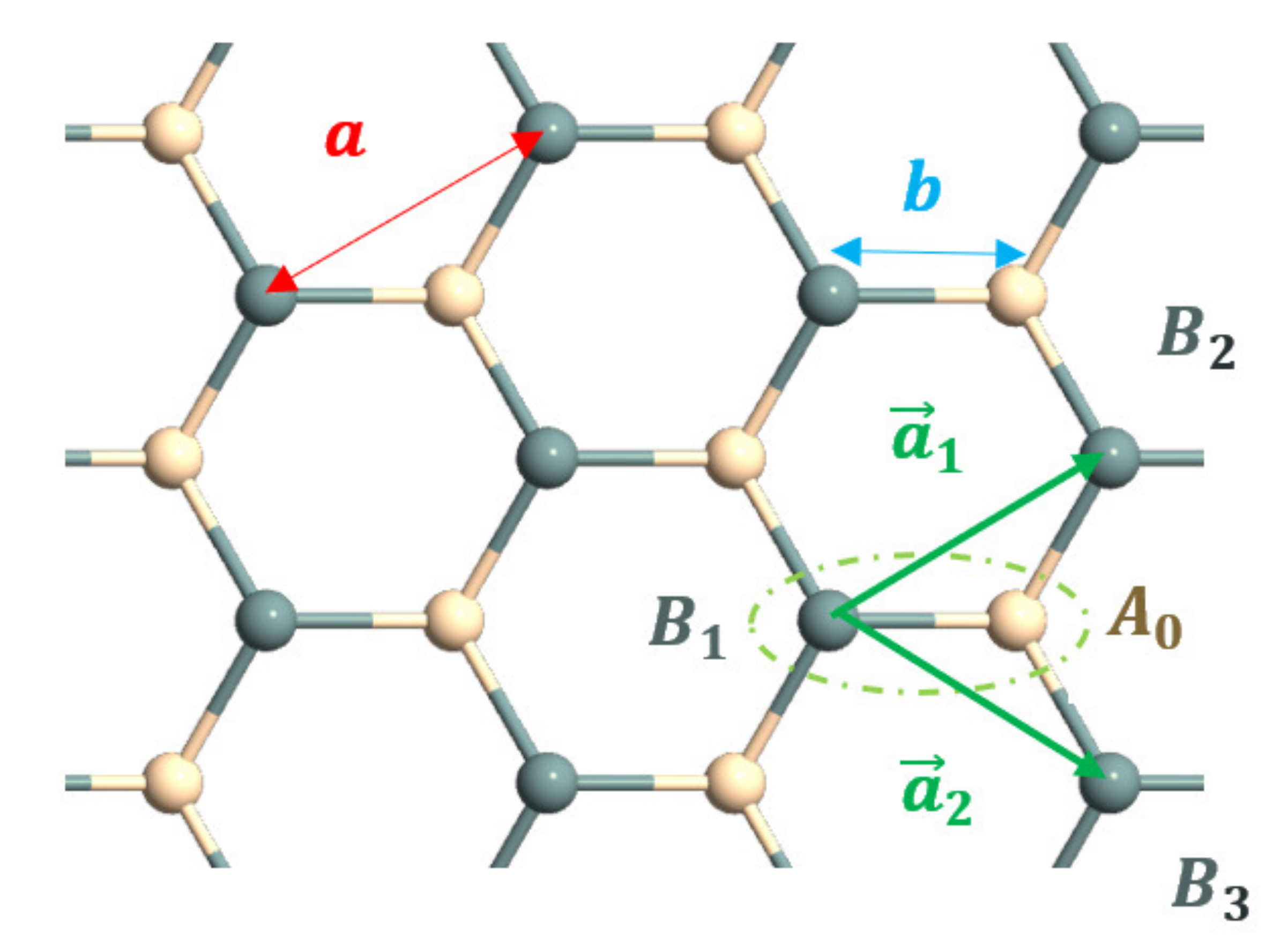}
				\includegraphics[width=1\textwidth]{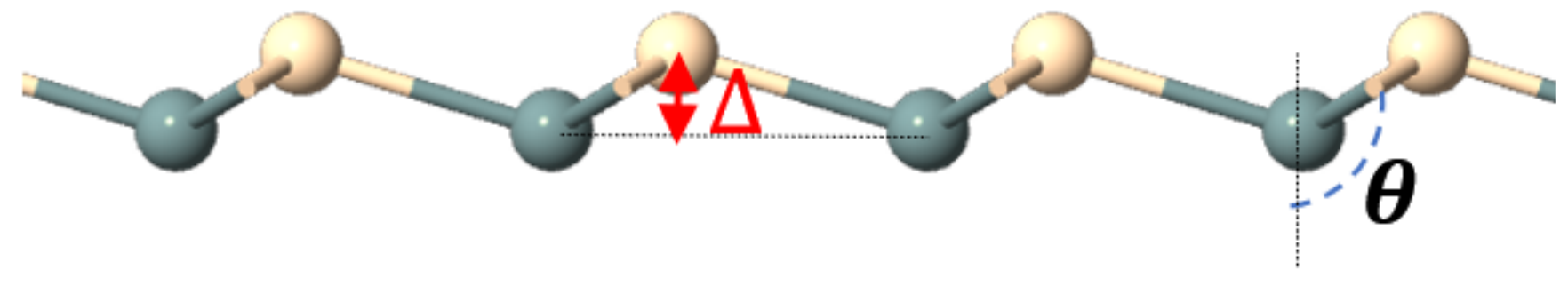}
			\end{minipage}
		\end{tabular}
		\caption{\label{fig:1} (top) Schematic view of 2D honeycomb lattice, (bellow) buckling parameter, $\Delta$, and Azimuth degree, $\theta$.}
	\end{figure} 
	
	The relaxation procedure of atomic structures of these compounds has been shown that almost a little more than half of these structures have been relaxed in a planar geometry and others have relaxed in low-buckled (LB) geometry. So, the 2D hexagonal structure of binary compounds of group-IV elements, their binary compounds, and group III-V compounds all form a honeycomb structure. 
	
	The 2D hexagonal lattice in which atoms have been arranged to form a PL honeycomb structure and LB one as shown in Fig.~\ref{fig:1}. There is a significant difference between PL and LB geometry that is the buckling parameters $\Delta$, causes covalent $\sigma$ bonds derived from the planar hybrid $sp^2$ orbitals between adjacent atoms, changed to $sp^3$ hybridization. The buckling parameter has been reported in Table.~\ref{tab:1} based on the lattice constant, $a$. Among these materials, germanene has a maximum value of $\Delta$ in terms of lattice constant, $a$, and other LB compounds have the values of $\Delta$ less than $20$ \textdiscount~ of lattice constant. Next, first nearest-neighbor distance (b), which will be used in \ref{sec:AppA}, is exactly equal to $\frac{a}{\sqrt{3}}$ in PL structures.
	
	\begin{table}[!h]
		\begin{center}
			\caption{\label{tab:1} Calculated results for group-IV elements, and group III-V compounds having a 2D honeycomb structure. Stable structures are identified as PL or LB standing for the planar and low-buckled geometries, respectively. The values of buckling parameters $\Delta$; 2D hexagonal lattice constant, $|\vec{a_1}|=|\vec{a_2}|=a$, are given. Some of the structural parameters are illustrated in Fig.~\ref{fig:1}.}
			\begin{tabular}{l l c c c c c c c c}
				\hline \hline
				& &$a$&$b$&$\Delta$&\multicolumn{2}{c}{band-gap$(eV)$}&$\theta$&$v_F$(FP)&TI \\
				& Geo. 	&$(\AA)$& $(\AA)$	& &no SOI&with SOI&degree &$\times 10^5$$(m/s)$&(init.) \\ \hline
				& Grp.		& III-V	&	&                           \\ 			
				BN      & PL   		& 2.51	&1.45& - &  4.66 &   4.66   &90.00 &0.102  &no  \\ 
				BP      & PL   		& 3.18 	&1.83& - &  0.88  &   0.88    &90.00 &0.053 &no  \\ 
				BAs     & PL   		& 3.36	&1.94& - &  0.76  &   0.76    &89.99 &0.014 &no  \\ 
				BSb     & PL   		& 3.69	&2.13& - &  0.37  &   0.36    &90.00 &0.050 &no  \\ 
				BBi     & LB   		& 3.82	&2.20& 0.12$a$  & 0.50&  0.47&101.39&0.058 &no  \\ 
				& 			& & 		&                           \\ 						
				& Grp.		& IV & 	&                                  \\  						
				Graphene& PL   		& 2.46	&1.42& -         & 0.00&0.00   &90.00 &8.31  &yes\\ 
				Silicene& LB   		& 3.80	&2.19& 0.19$a$  & 0.00&0.00   &108.42&5.06  &yes\\
				Germanene& LB   	& 3.93	&2.27& 0.20$a$  & 0.02 &0.02  &109.63&4.68  &yes\\
				Stanene & LB   		& 4.54	&2.62& 0.19$a$  &0.00& 0.07   &108.46&4.85  &yes\\
				SiC     & PL   		& 3.12	&1.80& -         &2.40&2.40    &90.00 &0.12 &no\\
				SiGe    & LB   		& 3.86	&2.23& 0.16$a$  &0.01 &0.02 &105.55&3.64  &yes\\
				GeC     & PL   		& 3.19	&1.84& -         &2.08&2.13    &89.96 &0.15 &no\\
				SnC     & PL   		& 3.61	&2.08& -         &1.69&1.68    &89.97 &0.11 &no\\
				SnSi    & LB   		& 4.21	&2.43& 0.17$a$  &0.25&0.23  &106.88&0.42 &no\\
				SnGe    & LB   		& 4.28	&2.47& 0.19$a$  &0.23&0.19  &107.89&0.49 &no\\
				\hline \hline
			\end{tabular}
		\end{center}
	\end{table}
	
	Some aspects of the stability of these materials such as phonon analysis had been studied in Ref.~\cite{Sahin_2009_IIIV}. We also report the carrier Fermi velocity ($v_F$) around the Dirac point K from first-principles with linear approximation, which is comparable with the results of Ref.~\cite{Liu_2011_EFHTB}.
	
	\section{\label{sec:ElS} Electronic structures}
	
\begin{figure}[!ht]
		\centering
		\begin{tabular}[c]{cc}
			\begin{tabular}[c]{cc}
				\begin{minipage}[c]{0.53\textwidth}
					\includegraphics[width=\textwidth]{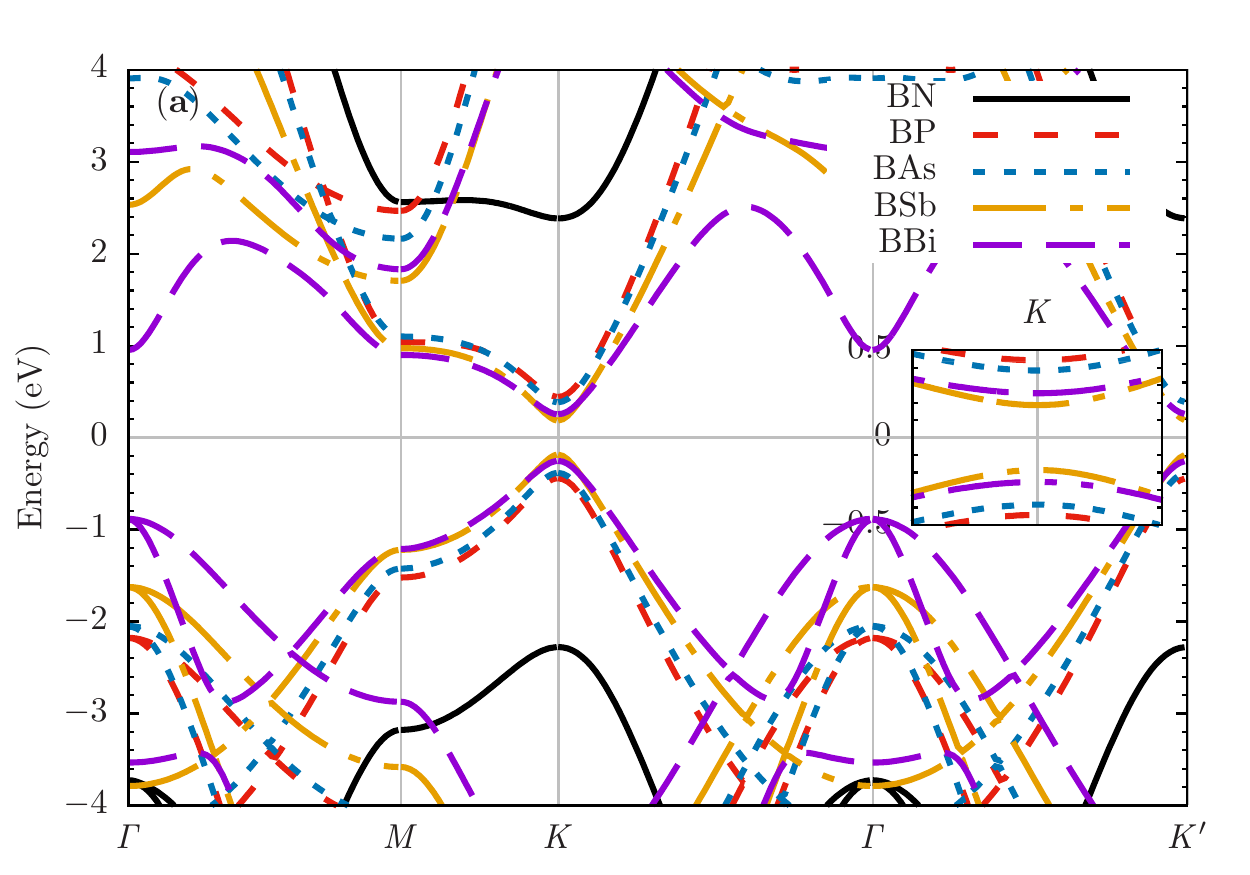}
				\end{minipage}\\
				\vspace{-5.0mm}
				\begin{tabular}[c]{cc}
					\begin{minipage}[c]{0.53\textwidth}
						\includegraphics[width=\textwidth]{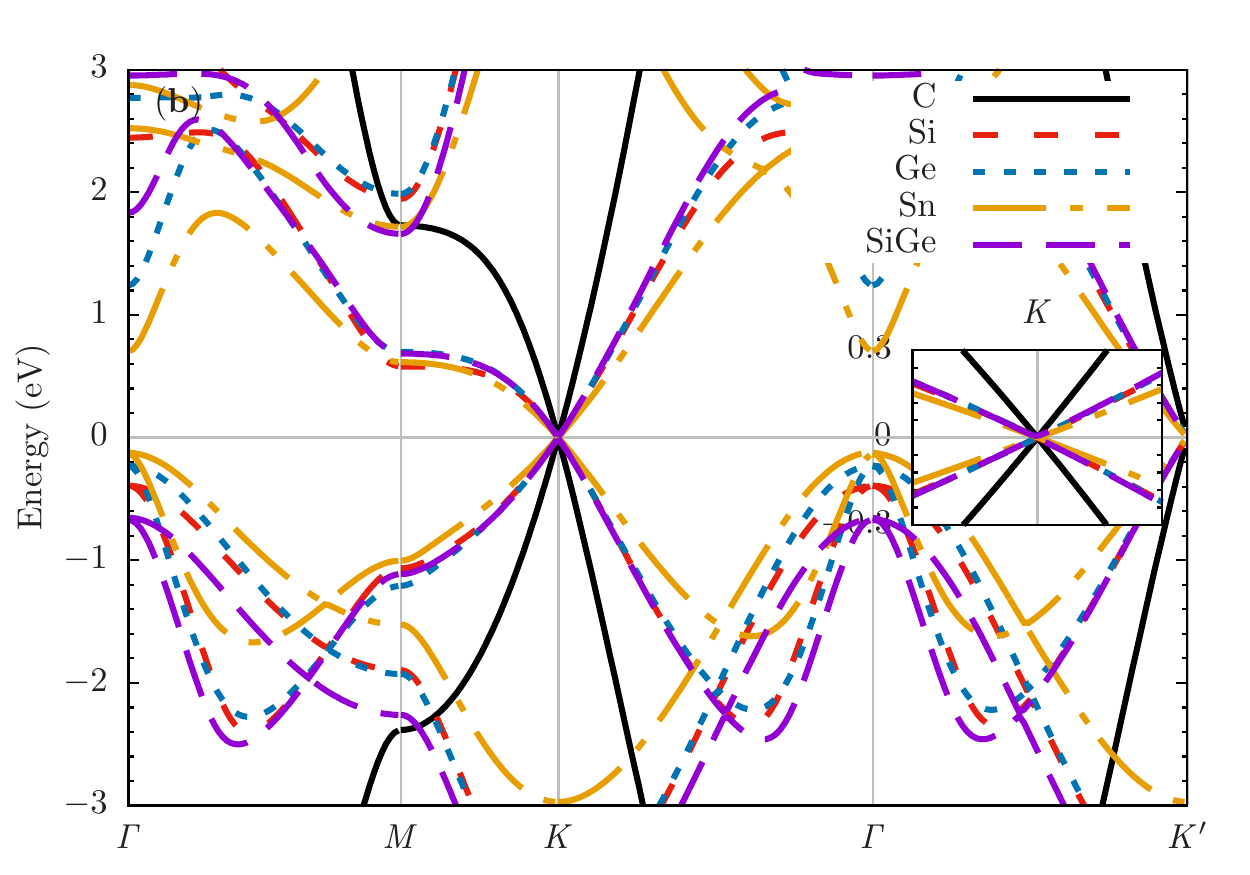}
					\end{minipage}					
					\hspace{-6.0mm}
					\begin{minipage}[c]{0.53\textwidth}
						\includegraphics[width=\textwidth]{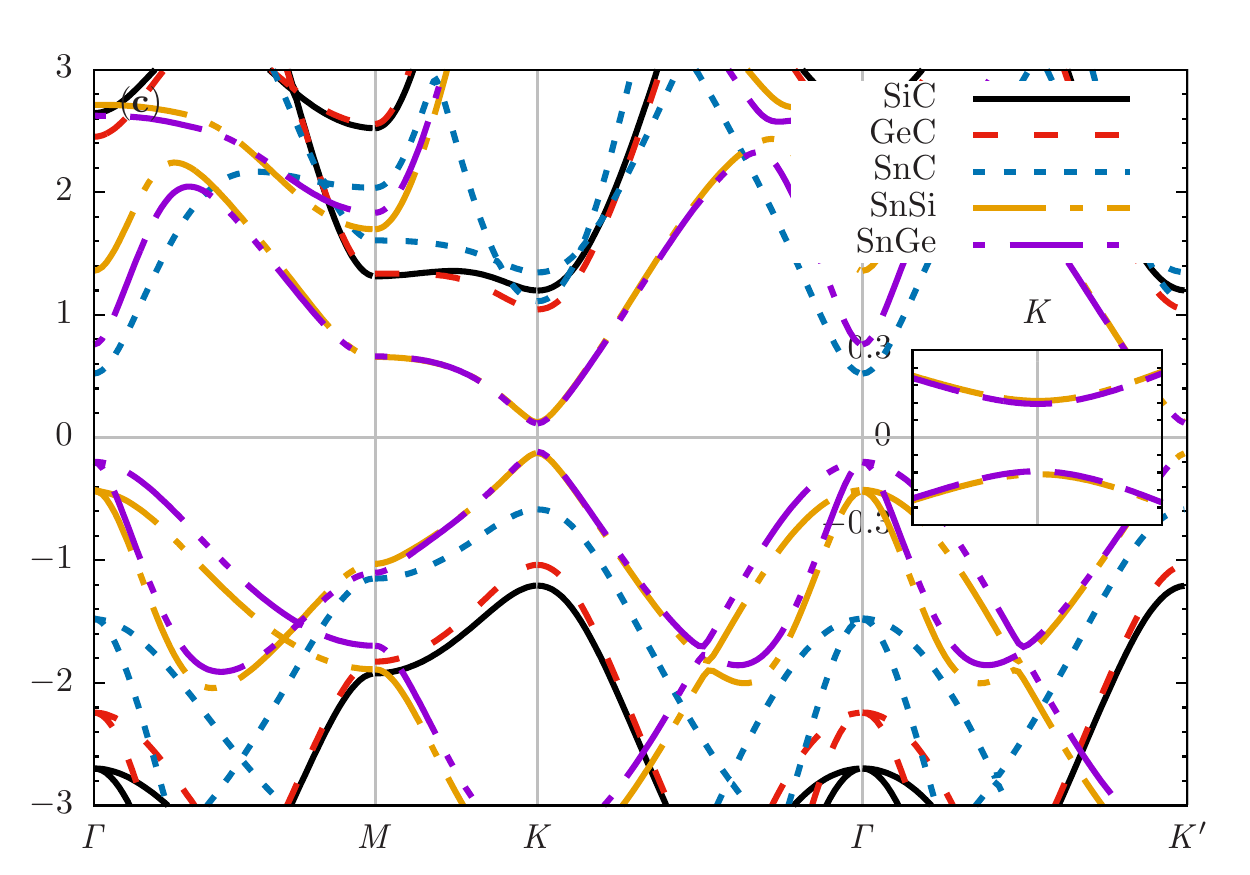}
					\end{minipage}
				\end{tabular}
			\end{tabular}
		\end{tabular}
		\caption{\label{fig:2}(color online). DFT band structure of monolayer of (a) BX (X=N, P, As, Sb, Bi), (b) group IV crossed (zero band-gap), and (c) group IV with non-zero band-gap; inset figures show the band near Fermi level around $K$ high symmetry point within a narrower energy window.}
	\end{figure}

	Band structures of group III-V compounds have been shown in Fig.~\ref{fig:2}(a), exhibited all electronic bands have a band-gap more than $0.4~eV$, but BSb 2D honeycomb lattice has the least band-gap, a little less than $0.4~eV$. The authors know that DFT calculation based on GGA approximation underestimates band-gap \cite{Borlido_2020_gapVSgga}, however, the pattern of change in this value is important for the appearance of the topological phase. In the group-IV materials (Fig.~\ref{fig:2}~(b),(c))) those which contain a Dirac point at $K$ point are like Graphene, and others have band-gap values from nearly $0.25~eV$ in SnSi and SnGe 2D monolayers to more than $2~eV$ in SiC, GeC, and SnC 2D honeycomb structures. Germanene has the highest buckled parameter, and like group IV-IV crossed structures, the Dirac point is exactly settled on the fermi level despite earlier report~\cite{Trivedi_2014_silicene_Germanene}. All of the band structures have been derived without considering SOI, then when we want to mention the topological phase, need to explore energy bands considering SOI.
	
\begin{figure}[!hb]
		\centering
		\begin{tabular}[c]{cc}
			\begin{tabular}[c]{cc}
				\begin{minipage}[c]{0.53\textwidth}
					\includegraphics[width=\textwidth]{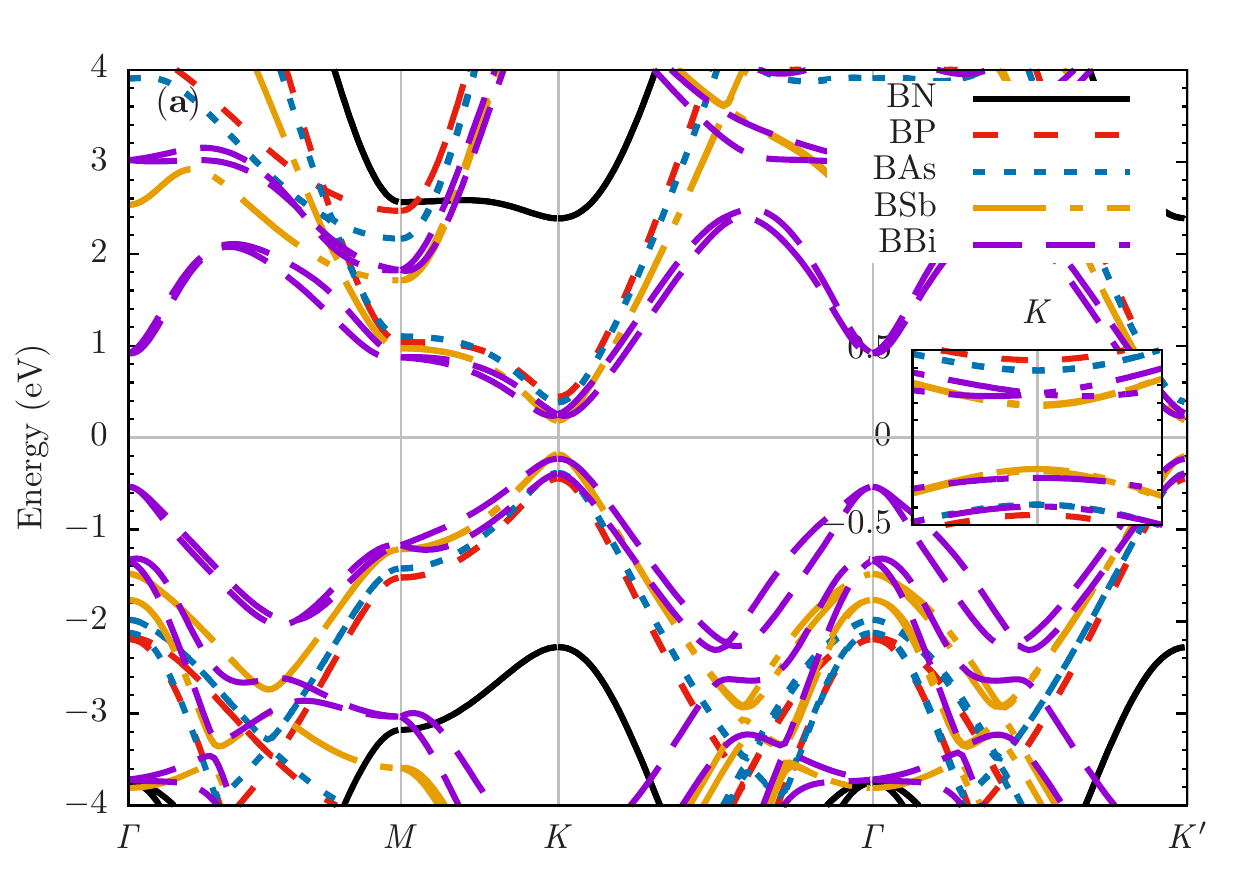}
				\end{minipage}\\
				\vspace{-5.0mm}
				\begin{tabular}[c]{cc}
					\begin{minipage}[c]{0.53\textwidth}
						\includegraphics[width=\textwidth]{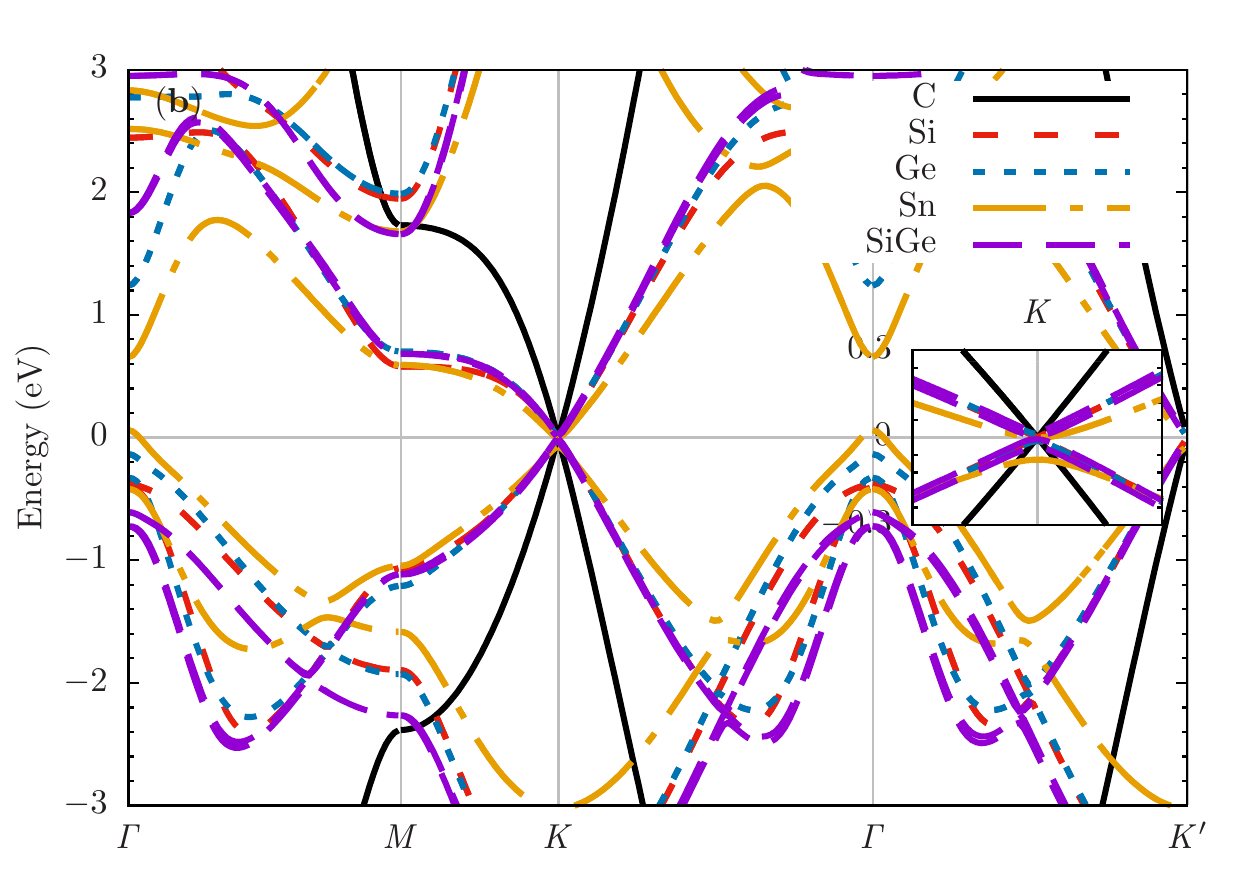}
					\end{minipage}					
					\hspace{-6.0mm}
					\begin{minipage}[c]{0.53\textwidth}
						\includegraphics[width=\textwidth]{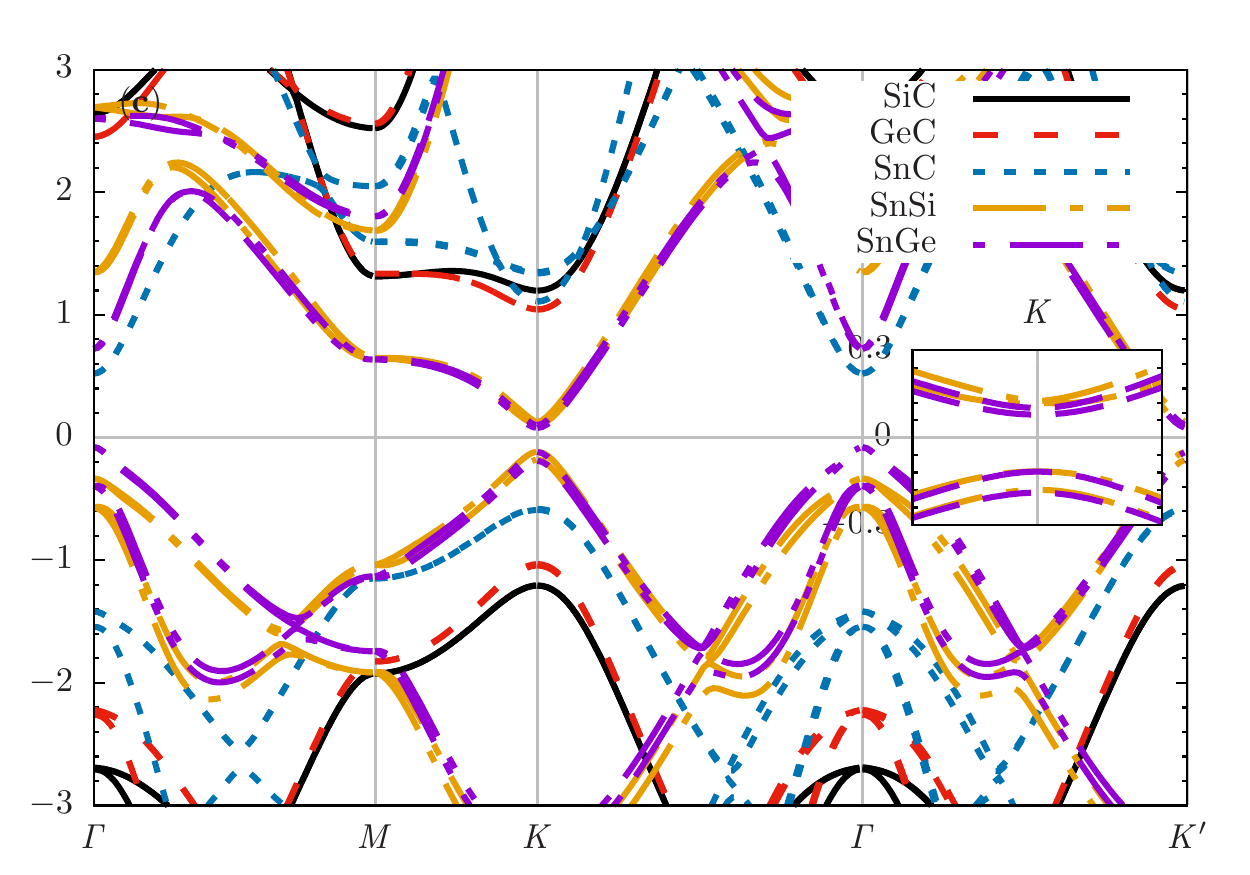}
					\end{minipage}
				\end{tabular}
			\end{tabular}
		\end{tabular}
		\caption{\label{fig:3}(color online). DFT band structure in presence of SOI for a monolayer of (a) BX (X=N, P, As, Sb, Bi), (b) group-IV crossed (zero band-gap), and (c) group-IV with non-zero band-gap; inset figures show the band near Fermi level around K high symmetry point within a narrower energy window.}
	\end{figure}

	When the calculations include SOI, band structures modify in the compounds with heavier elements, like antimony (Sb), bismuth (Bi) as shown at Fig.~\ref{fig:3}(a), and tin (Sn) has been exhibited at Fig.~\ref{fig:3}(b), (c). As we have expected SOI is well strong in BSb and BBi honeycomb structures, therefore, in the BX (X=N, P, As, Sb, Bi) honeycomb lattices, these structures are candidates for TI phase. So, we selected BSb monolayer structure, especially due to PL geometry relative to another candidate, so the discussion about Rashba SOI will be omitted \cite{Liu_2011_EFHTB}. Similarly, In the group III-V compounds, the monolayer of stanene and SiGe seems more appropriate due to larger SOI effects.
	
	As the Fig.~\ref{fig:3} shows, the effective SOI in LB geometry opens a band-gap at the Dirac points and establishes the QSH effect. The values of band-gap with and without SOI summarize in Table~\ref{tab:1}. We refer to this discussion later.
	
	\section{\label{sec:Tpw} Topological view point via first-principles calculations}
	
	In 2D structures, by changing some physical parameters such as external EF perpendicular to the lattice surface, and SOI strength, a Dirac point appear near the Fermi surface \cite{Ast_2012_efield}, and we expect a non-trivial topological phase. In the following, we consider some structures from Table~\ref{tab:1} based on the ability to close the band-gap under changing these physical parameters.
	
	Now, we try to classify some candidates with narrower band-gap (lower than 0.5 eV) for TI phase. There are some nominees such as BSb, SiGe, SnSi, and SnGe. In the following, we will examine the energy band structure in the presence of spin-orbit interaction and electric field.
	
\begin{figure}[!h]
		\centering
		\begin{tabular}[c]{cc}
			\begin{tabular}[c]{cc}
				\begin{minipage}[c]{0.50\textwidth}
					\includegraphics[width=\textwidth]{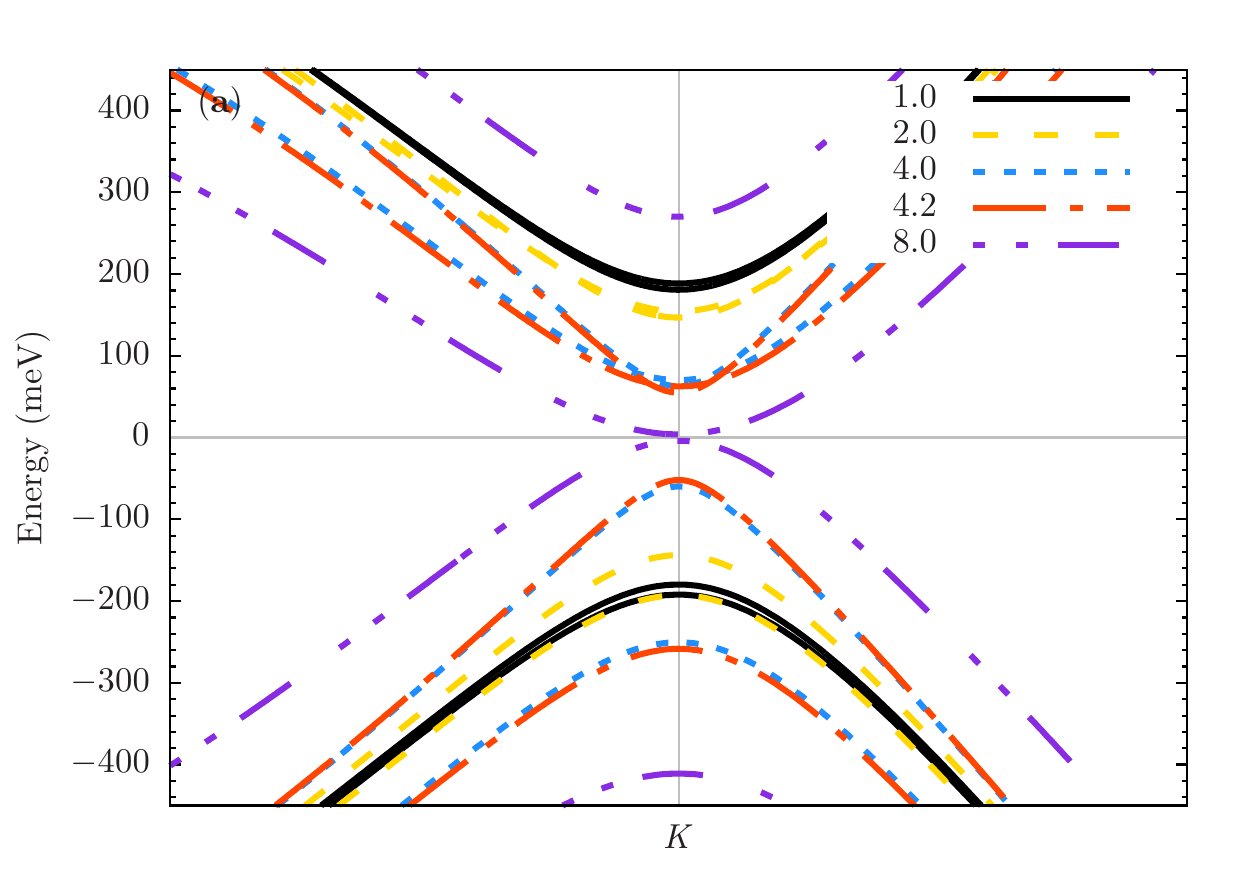}
				\end{minipage}\\
				\vspace{-3.0mm}
				\begin{tabular}[c]{cc}
					\begin{minipage}[c]{0.50\textwidth}
						\includegraphics[width=\textwidth]{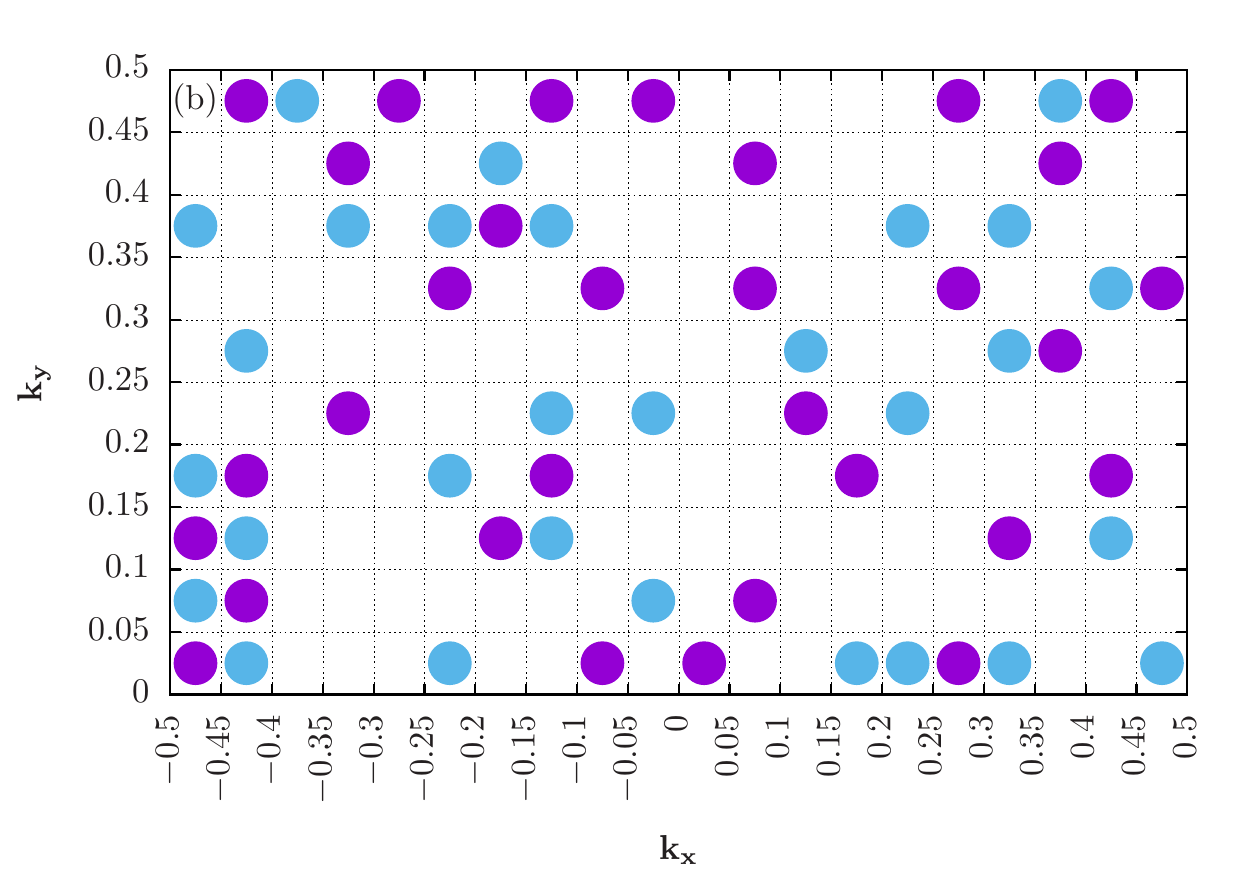}
					\end{minipage}
					\hspace{-6.0mm}
					\begin{minipage}[c]{0.50\textwidth}
						\includegraphics[width=\textwidth]{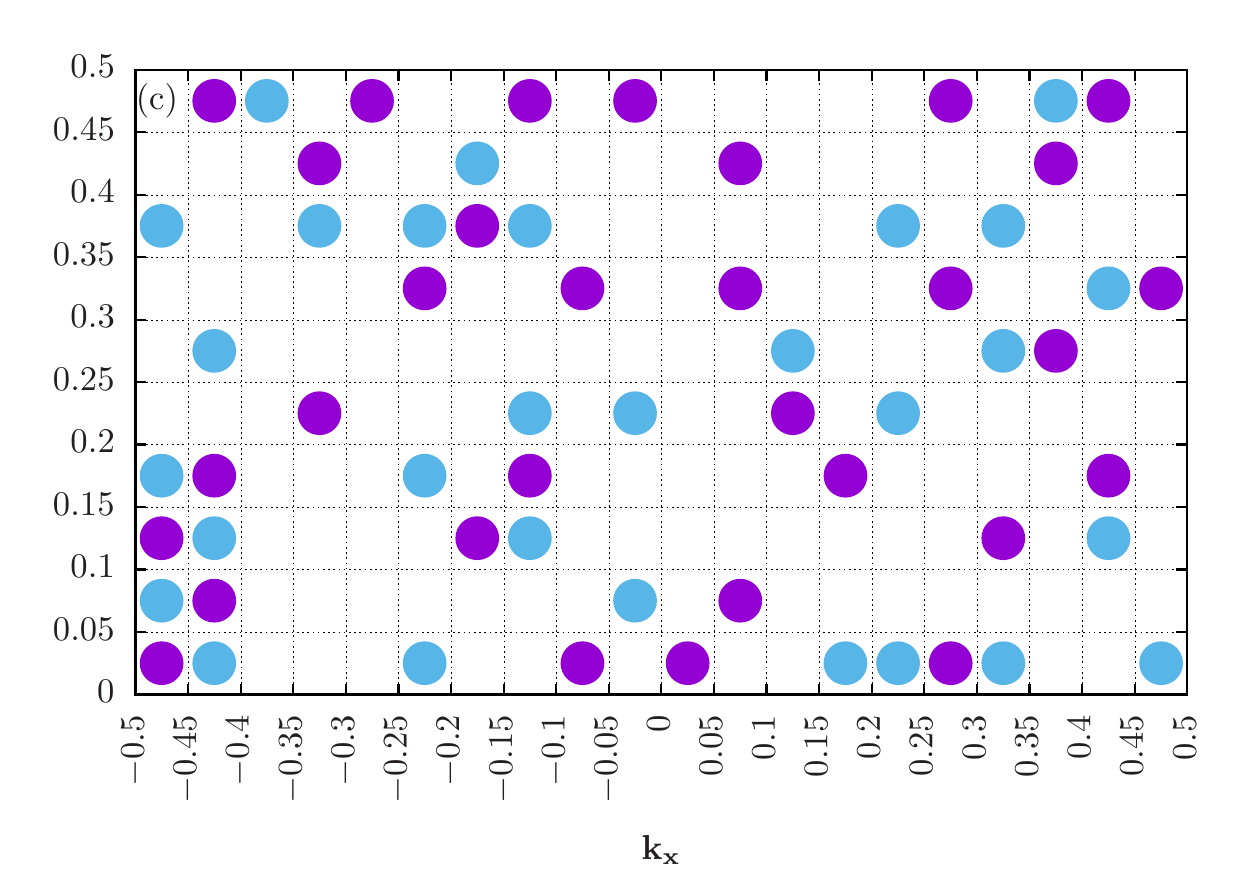}
					\end{minipage}
				\end{tabular}
			\end{tabular}
		\end{tabular}
		\caption{\label{fig:6}(color online). (a) DFT band structure for a monolayer of BSb in terms of the coefficients of the initial strength of SOI (the numbers represent multiples of initial SOI):  1.0 (solid, black), 2.0 (yellow, dashed), 4.0 (blue, dotted), 4.2 (red, dashed-dot-dashed), 8.0 (magenta, dot-dot-dashed). (b), (c) Calculated LCNs in discretized Brillouin zone. Violet (dark) circles and blue (light) circles indicate LCNs of +1 and -1, respectively. The $\mathbb{Z}_2$ topological invariant is computed as the total LCNs modulo two in the half Brillouin zone \cite{Sawahata_2018_Bi}, $\mathbb{Z}_2$ invariant is 0 and 1 for SOI=1 and SOI=4.2, respectively. }
	\end{figure}

	In the case of the monolayer of BSb when the strength of SOI changes, topological effects influence the band structure, and the band-gap tends to close, so trivial insulator changes to TI. Computing $\mathbb{Z}_2$ invariant, by using LCNs, as shown at Fig.~\ref{fig:6}~(b), (c), confirms that a phase transition occurred near a critical SOI strength. As this is depicted, the band-gap descends down slowly in $4.2$ times the initial strength of SOI, Table~\ref{tab:4} gives the band-gap values. In this table, for instance, $\times2$ means double in initial SOI value, and vice versa. As mentioned earlier in Sec.\ref{sec:Intro}, SOI can be indicated in two forms that Rashba SOI can be controlled by tuning a perpendicular electric field; so it will be reasonable to examine the general foundation of the effect of SOI change, in order to study of materials that show the same behavior as our example exhibited here.
	
	\begin{table}[!h]
		\begin{center}
			\caption{\label{tab:4} $\mathbb{Z}_2$ invariant and band-gap values by DFT for monolayer of BSb in terms of the coefficients of the initial strength of SOI (the numbers represent multiples of initial SOI).}
			\begin{tabular}{ccccccccc}
				\hline \hline
				$\frac{SOI}{initial~SOI}$ &1.0&2.0&4.0&4.2&8.0 \\ \hline
				$\epsilon_g (eV)$  &0.36&0.29&0.12&0.10&0.00      \\ 
				$\mathbb{Z}_2$ invariant  &0&0&0&1&1      \\
				\hline \hline
			\end{tabular}
		\end{center}
	\end{table}
	
\begin{figure}[!h]
		\centering
		\begin{tabular}[b]{ccc}
			\begin{minipage}[b]{0.55\textwidth}
				\includegraphics[width=\textwidth]{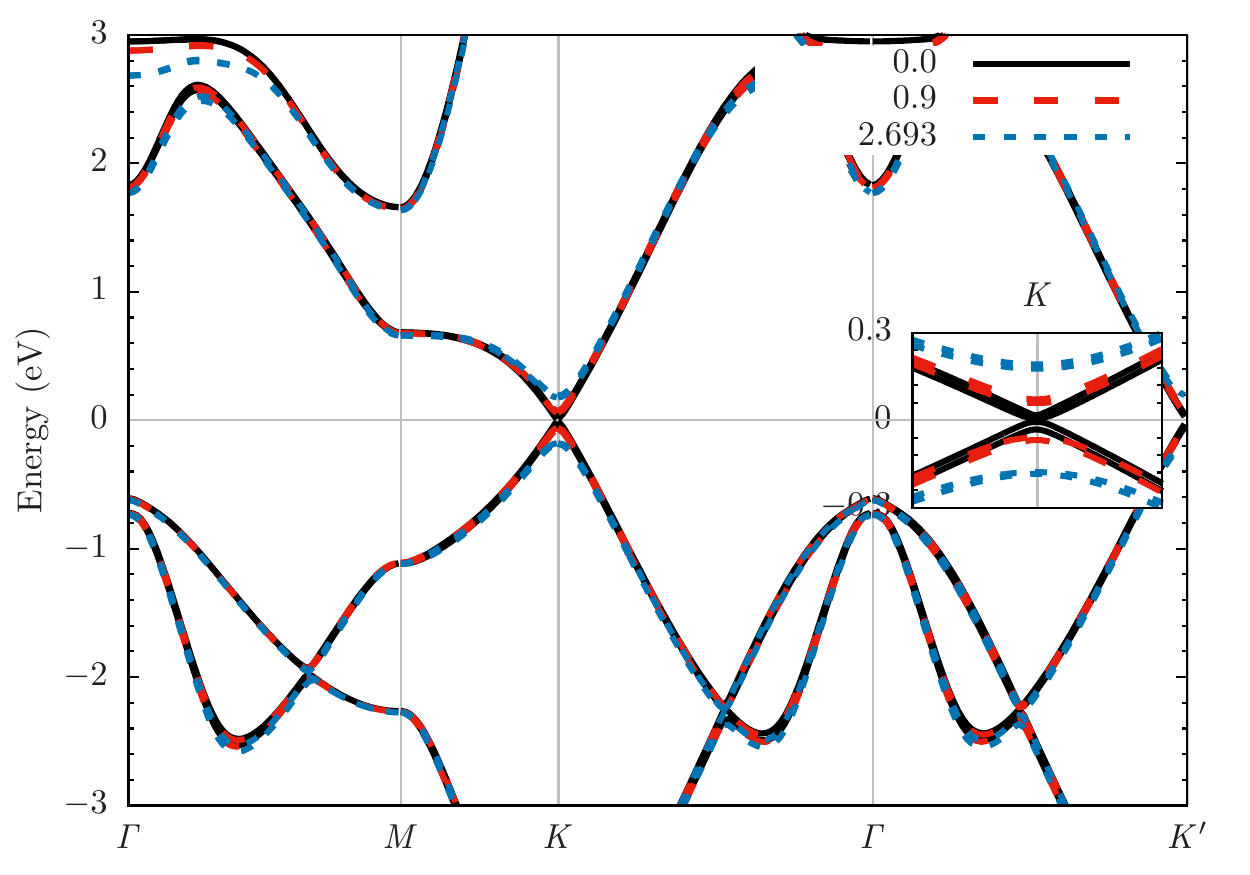}
			\end{minipage}
		\end{tabular}
		\vspace{-5.0mm}
		\caption{\label{fig:7}(color online). Band structures of the monolayer of SiGe by tunning EF strength, 0.0, 0.9 and 2.69 ($\frac{V}{\AA{}}$). }
	\end{figure}

	Next, as the Fig.~\ref{fig:7} shows for monolayer of SiGe, changing the EF strength perpendicular to the lattice plane, along the normal vector causes to disappear the Dirac cone at a distinctive EF. $\mathbb{Z}_2$ invariant calculation, as the total calculated LCNs modulo two in the half Brillouin zone, indicates that at nearly $2.7~\frac{V}{\AA{}}$ of EF, SiGe monolayer enters the trivial phase insulator, and the band-gap open dramatically as reported in Table~\ref{tab:5}.
	
	The intensity of the electric field used in these calculations is significant because its changes indicate a topological phase transition, as an example is reported by Sawahata et al.~\cite{Sawahata_2018_Bi}. Clearly, this pattern is reproducible for other materials with logical values of electric field strength.
	
	\begin{table}[!h]
		\begin{center}
			\caption{\label{tab:5} band-gap $(eV)$ Vs. EF strength ($\frac{V}{\AA{}}$) for SiGe monolayer.}
			\vspace{-5.0mm}
			\begin{tabular}{ccccccccccccccccc}
				\hline \hline
				$EF$&($\frac{V}{\AA{}}$)&-2.5&-1.2&-0.9&-0.6&-0.3&0.0&0.1&0.3&0.6&0.9&1.2&2.7 \\ \hline
				$\epsilon_{g}$&DFT&0.30&0.13&0.09&0.05&0.01&0.01&0.02&0.04&0.08&0.11&0.15&0.35	       \\
				($eV$)&TBA&0.08&0.06&0.06&0.06&0.06&0.07&0.07&0.08&0.09&0.11&0.14&0.35	       \\
				\hline \hline
			\end{tabular}
		\end{center}
	\end{table}
	
	\section{\label{sec:TB} Tight-binding approximation}
	
	In this section, with fitting the TB model and the DFT results, the Slater-Koster parameters for 15, 2D honeycomb compounds of group-IV and group III-V are calculated. Then, by considering the spin-orbit interaction and applying the electric field, topological features, concluded in the previous section, reproduce.
	
	\subsection{\label{sec:TB1} Tight binding parameterization from ab-initio calculations}
	
\begin{figure}[!h]
		\begin{tabular}[b]{cc}
			
			\begin{minipage}[b]{0.5\textwidth}
				\includegraphics[width=\textwidth]{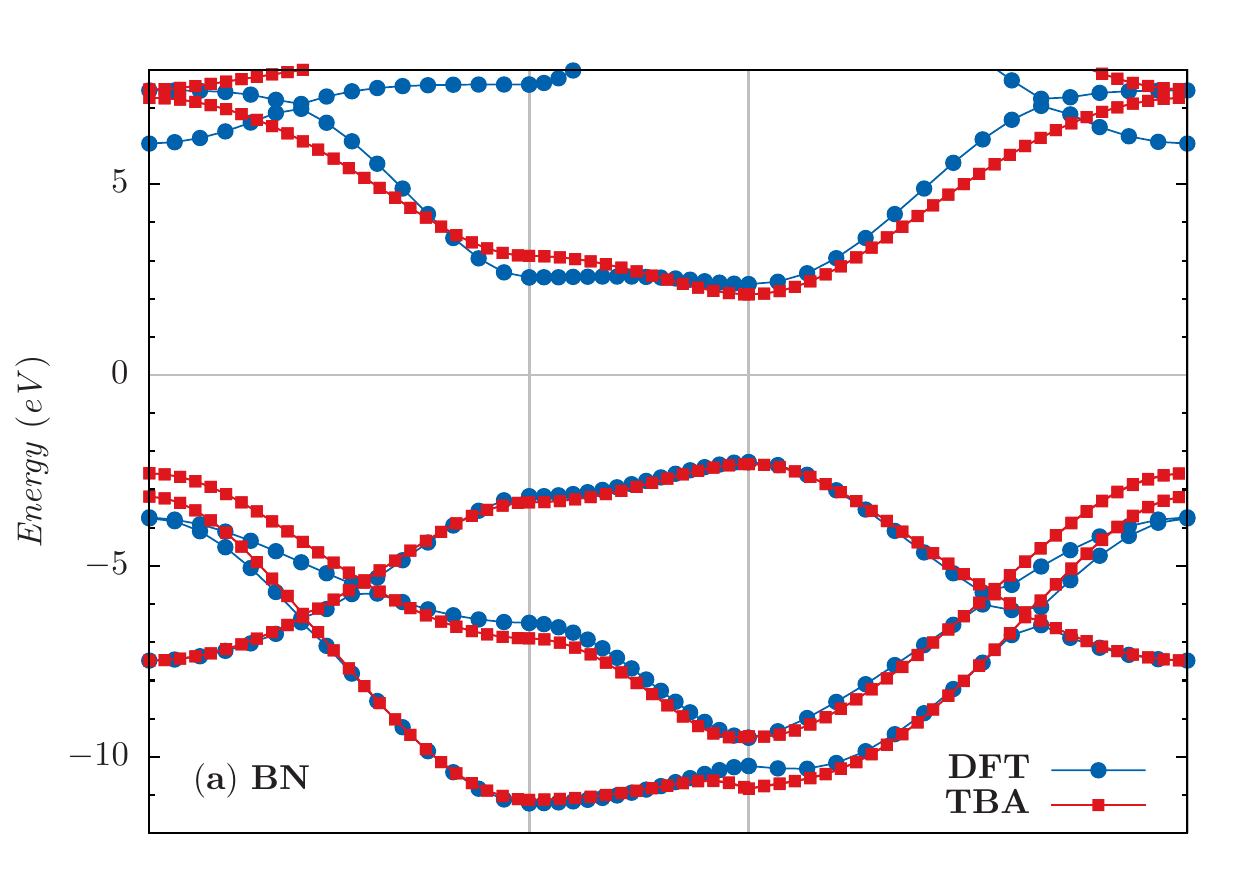}
				
				\includegraphics[width=\textwidth]{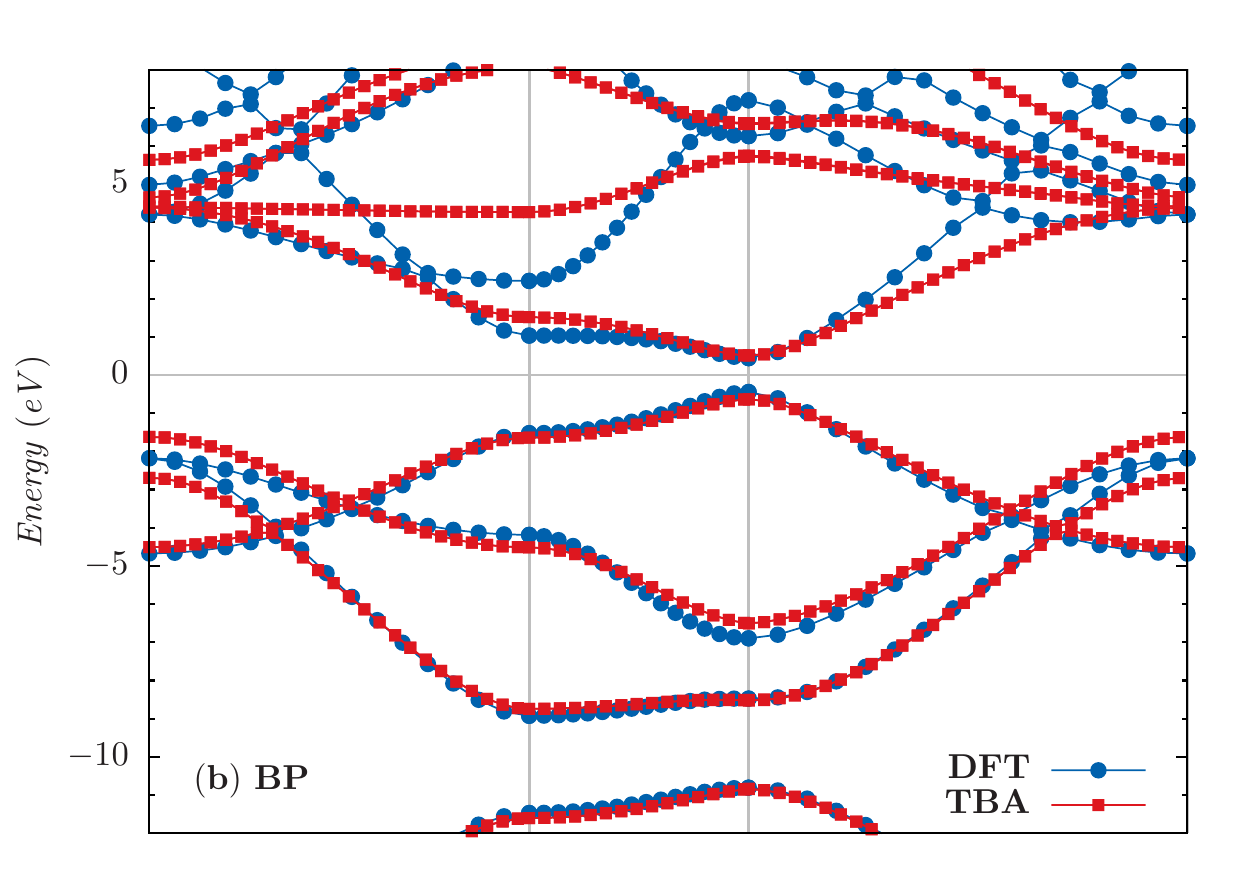}
				
				\includegraphics[width=\textwidth]{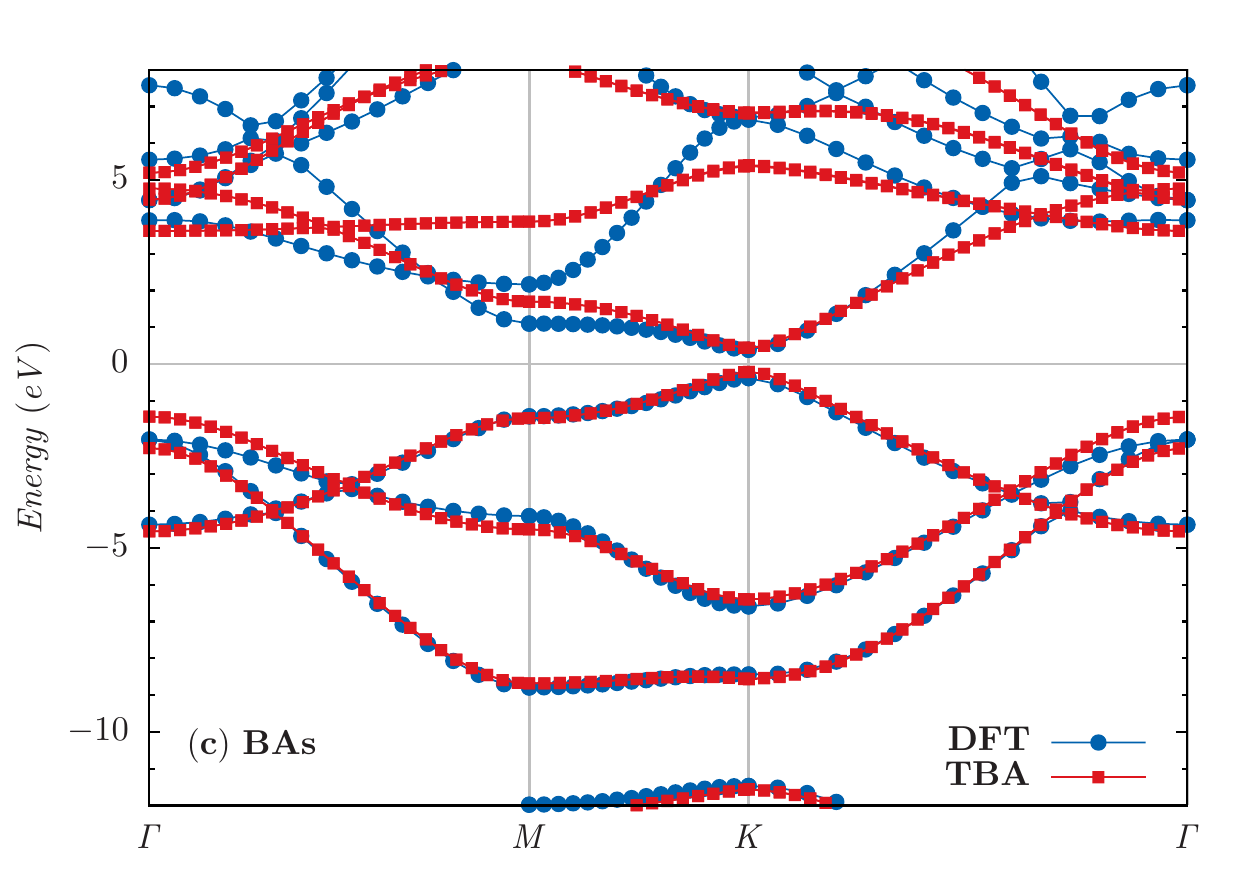}
			\end{minipage}
			\hspace{-7.40mm} 
			\begin{minipage}[b]{0.5\textwidth}
				\includegraphics[width=\textwidth]{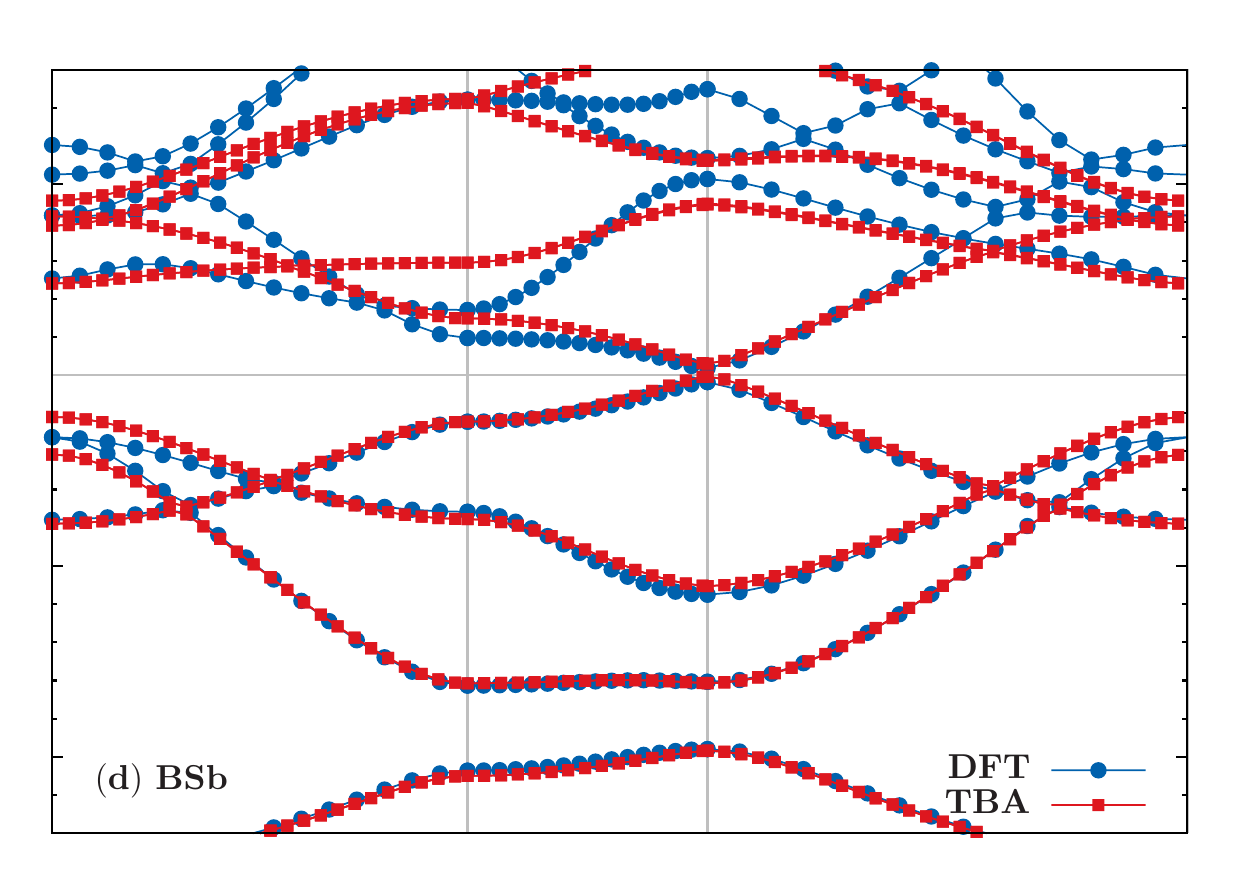}
				
				\includegraphics[width=\textwidth]{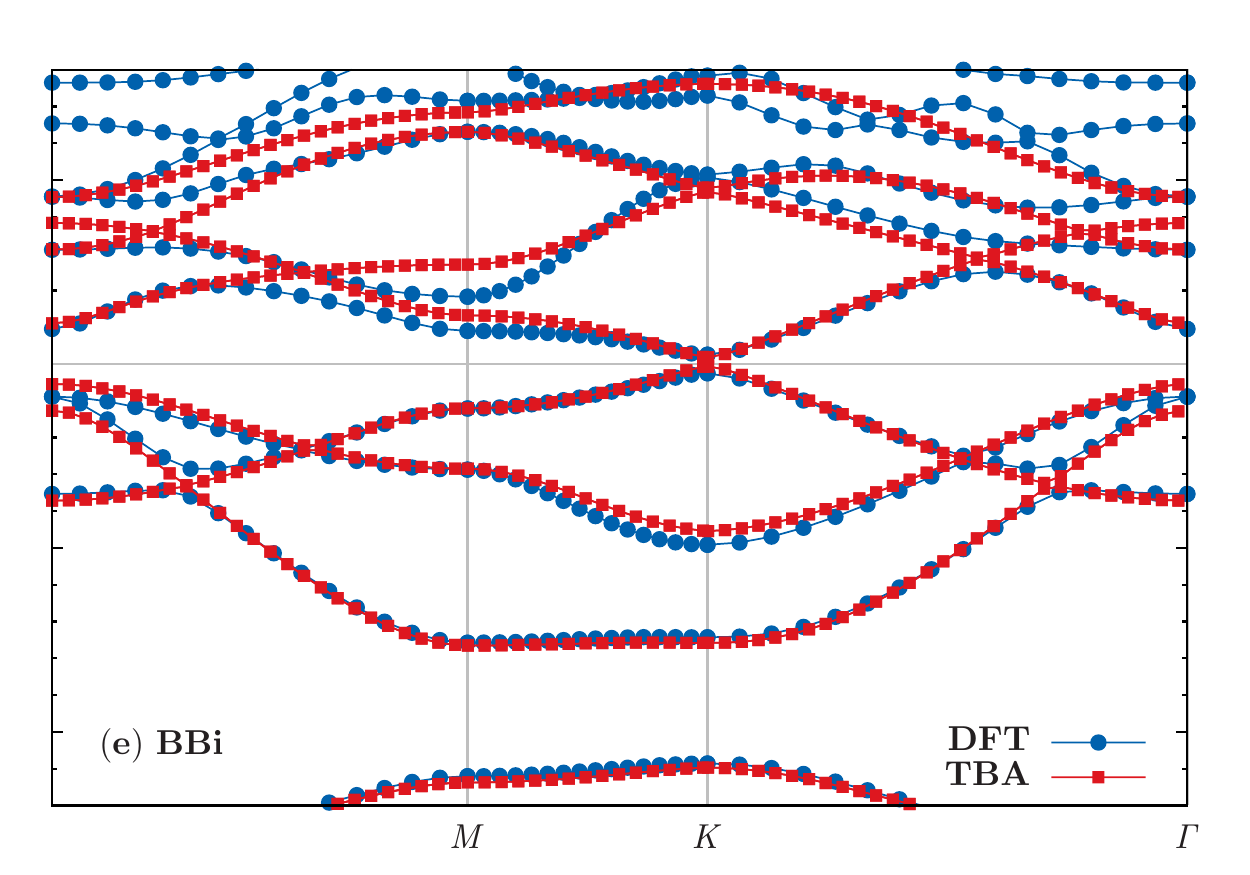}
			\end{minipage}
			
		\end{tabular}
		\vspace{-5.0mm}
		\caption{\label{fig:4}(color online). DFT Vs. TBA band structure for monolayer of (a) BN, (b) BP, (c) BAs, (d) BSb, (e) BBi; DFT bands plotted in light, circle (blue); and TBA bands plotted in dark, rectangle (red) points.}
	\end{figure}

\begin{figure}[!h]
		\begin{tabular}[b]{cc}	
			\begin{minipage}[b]{0.5\textwidth}
				\includegraphics[width=\textwidth]{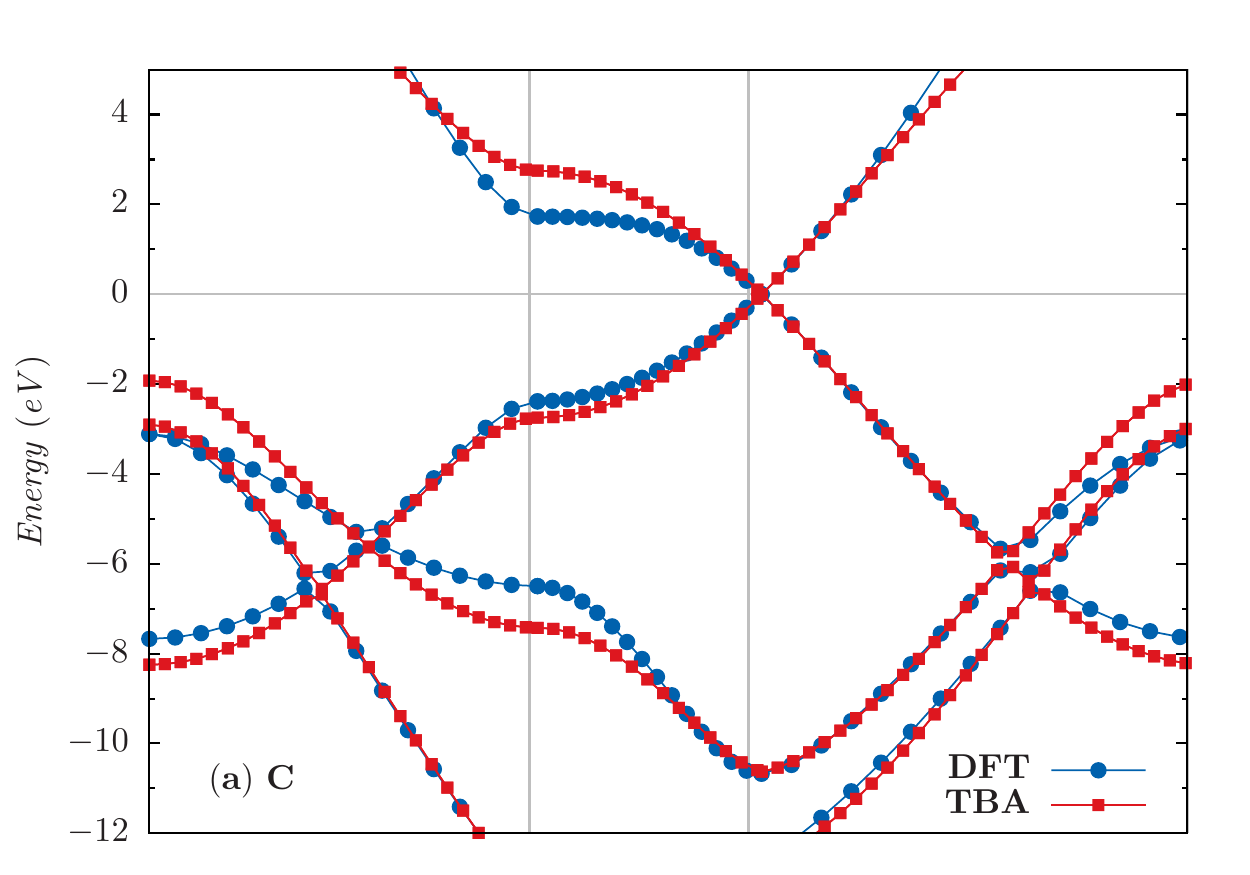}
				
				\includegraphics[width=\textwidth]{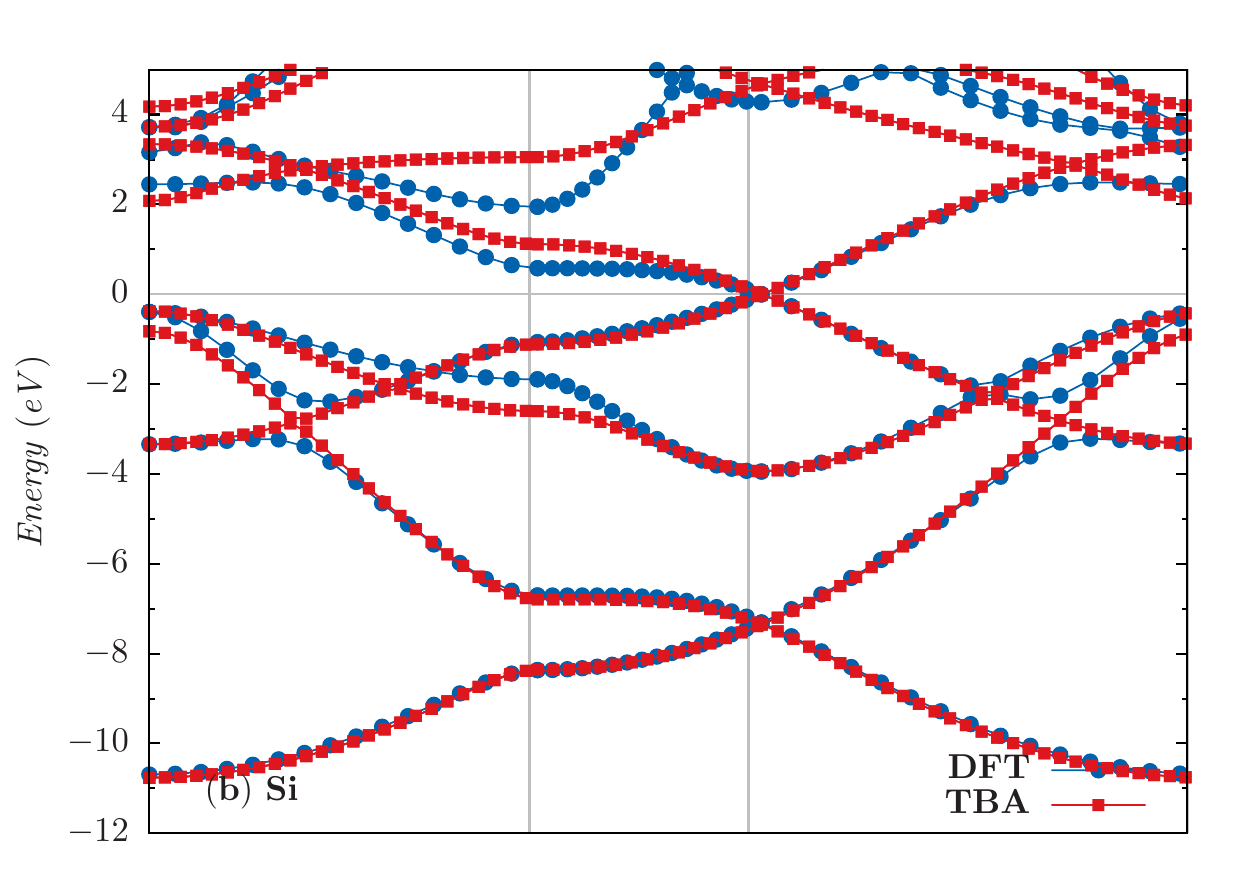}
				
				\includegraphics[width=\textwidth]{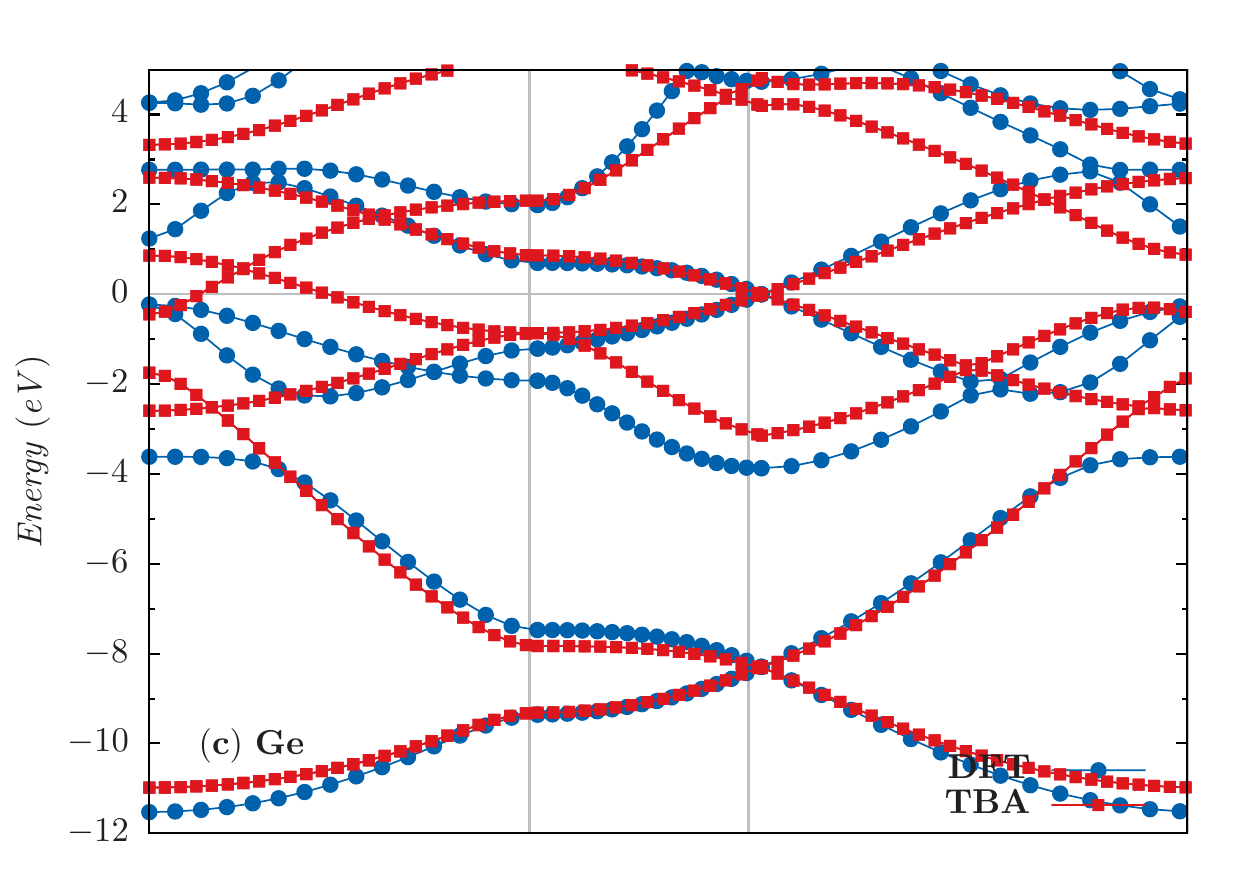}
				
				\includegraphics[width=\textwidth]{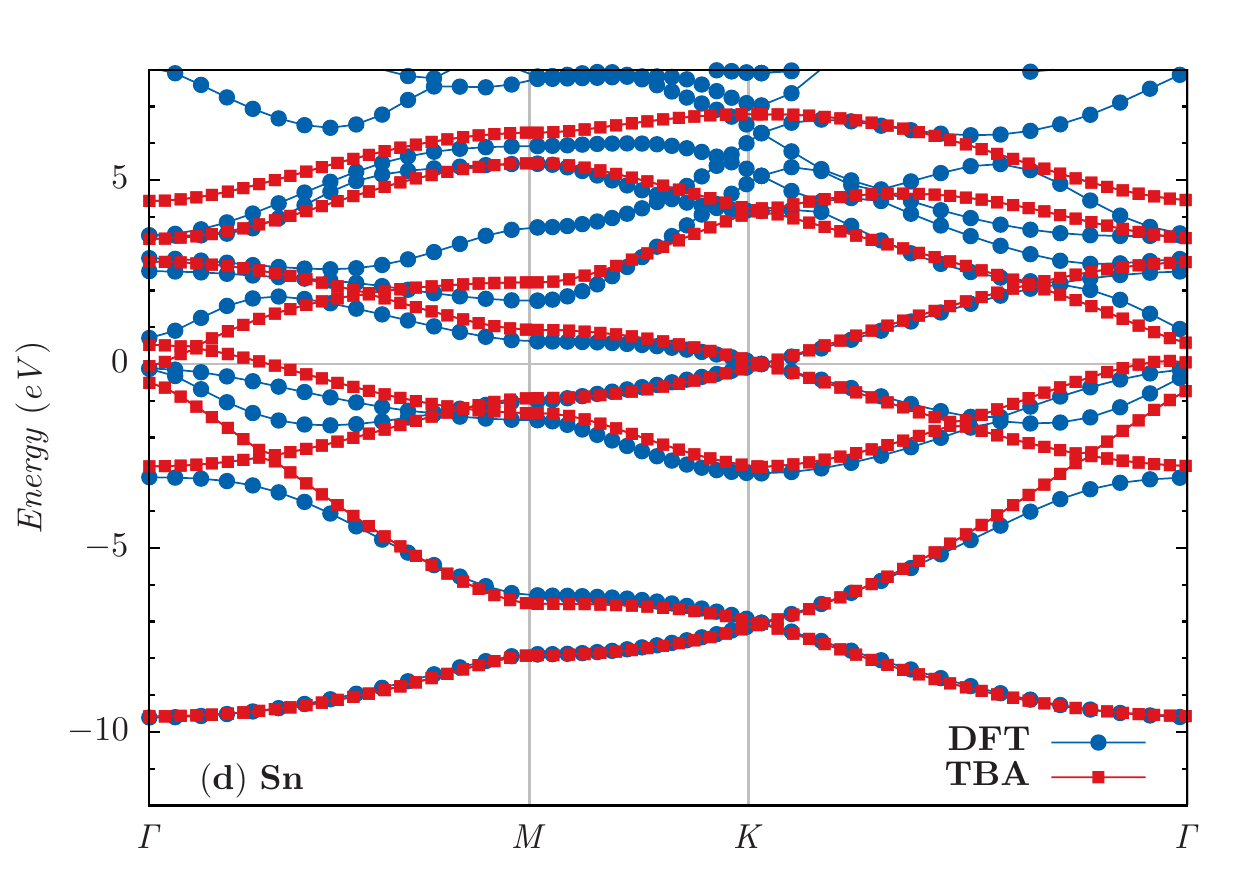}
			\end{minipage}
			\hspace{-7.40mm} 
			\begin{minipage}[b]{0.5\textwidth}
				\includegraphics[width=\textwidth]{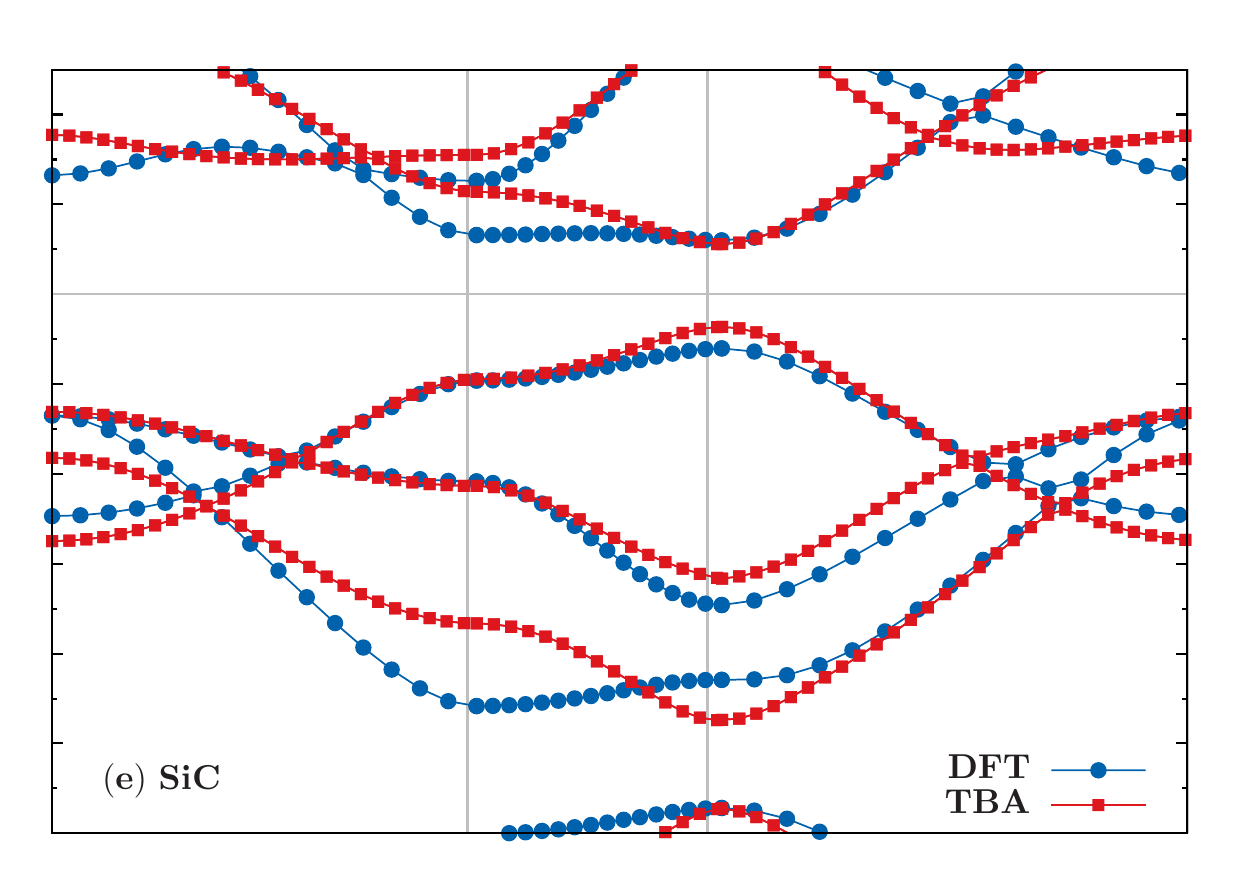}
				
				\includegraphics[width=\textwidth]{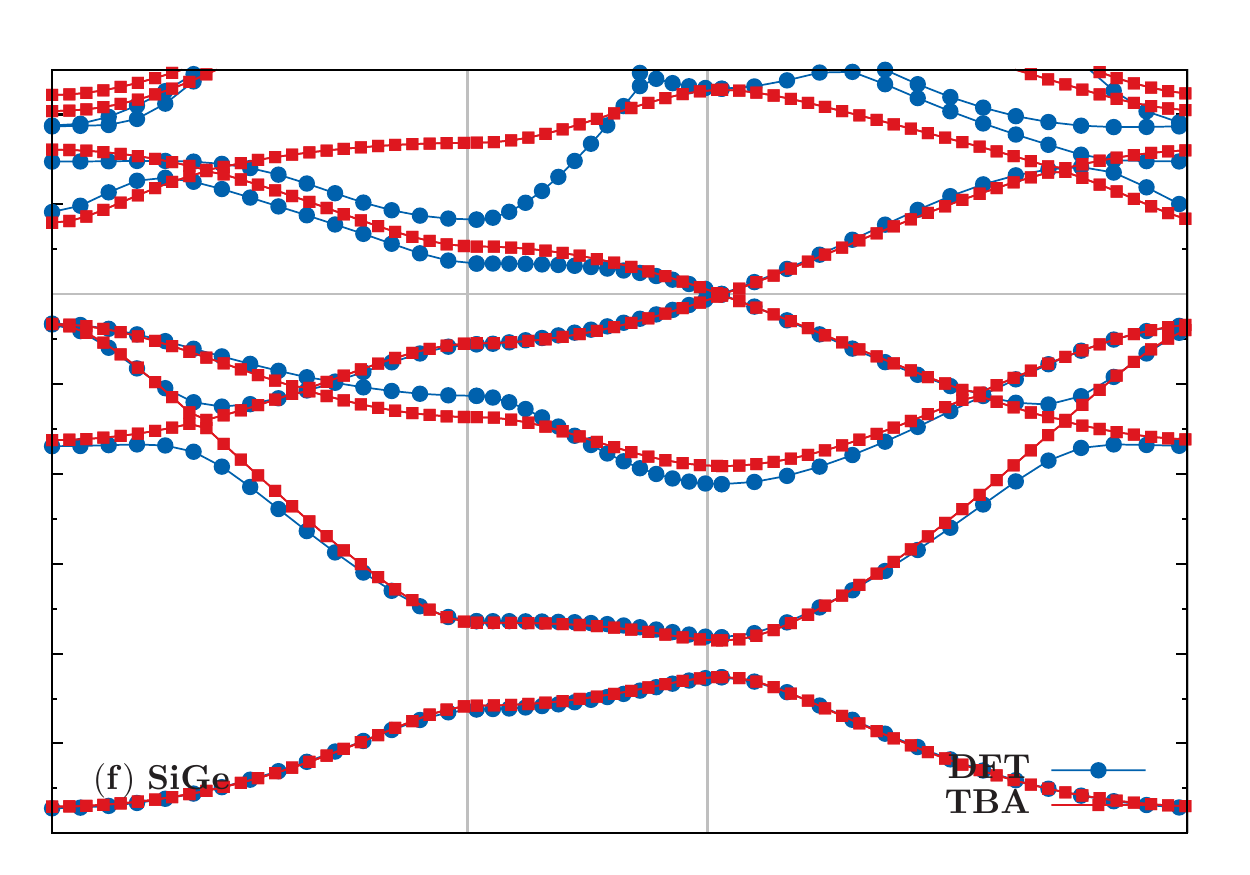}
				
				\includegraphics[width=\textwidth]{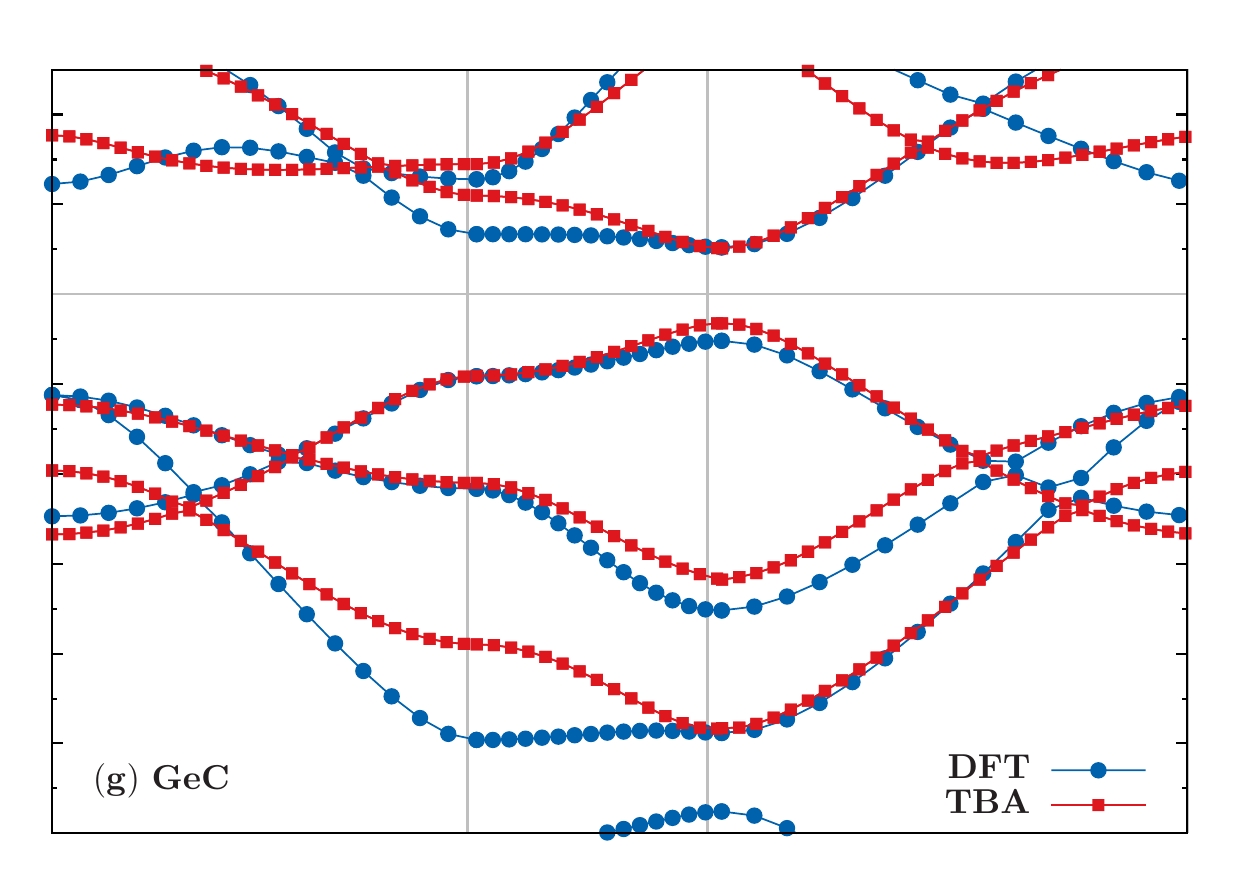}
				
				\includegraphics[width=\textwidth]{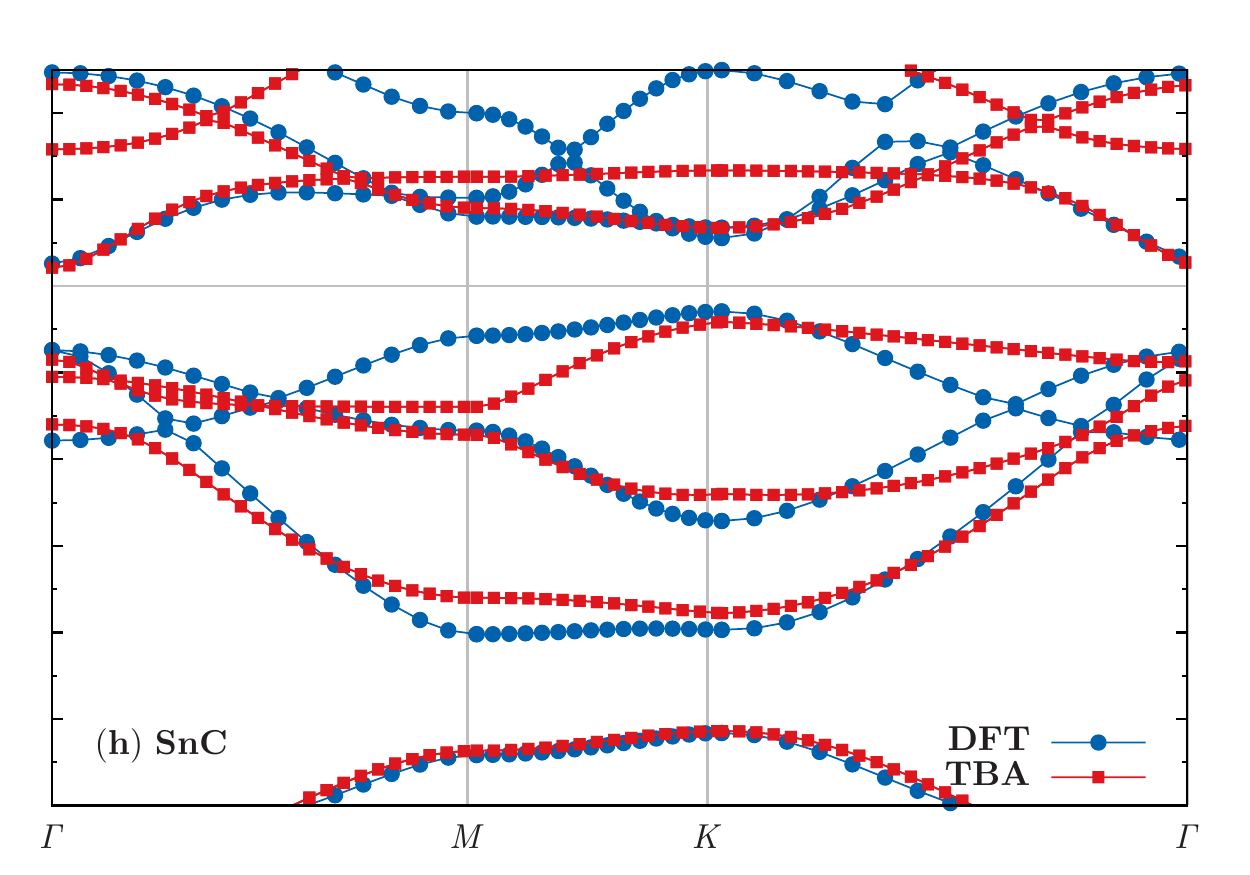}
			\end{minipage}
		\end{tabular}
		\vspace{-5.0mm}
		\caption{\label{fig:5}(color online).  DFT Vs. TBA band structure for (a) graphene, (b) silicene, (c) germanene, (d) stanene, and monolayer of (e) SiC, (f) SiGe, (g) GeC, (h) SnC; DFT plotted in light, circle (blue); and TBA plotted in dark, rectangle (red) points.}
	\end{figure}

	Based on the review will be denoted in \ref{sec:AppA}, we have used the $sp^3$ TBA to conclude the band structures. The result is illustrated in Fig.~\ref{fig:4} for group III-V compounds. As the plots show there is a good agreement with ab-initio calculation. The used parameters have been summarized in Table~\ref{tab:2}, and the results for group-IV compounds have been shown at Fig.~\ref{fig:5}, and the used parameters have been summarized in Table~\ref{tab:3}.
	
	When a structure has a preference towards LB than PL, $sp^2$ hybridization changes to $sp^3$. In PL mode the lobes of each atom are perpendicular to the plane of the slab and this leads to the information of $\pi$ bonds with nearest neighbors leading to conducting nature of the slab, but in LB structures the lobes of neighboring atoms point in opposite directions so the $\pi$ bands can only be formed with the second nearest neighbor than the first neighbor. The $sp^2$ hybridized orbitals get slightly hybridized into $sp^3$-like orbitals which causes weakening of $\pi$ bonds leading to buckled structure of materials like silicene, germanene, stanene, and the monolayer of SiGe, SnSi and SnGe; and BBi.
	
	In this paper, for the purpose of simplicity and more efficiency, we have just tried to derive the DFT bands considering only with first nearest neighbors. Table~\ref{tab:2}, and Table~\ref{tab:3} have been summarized all the parameters to obtain the band structures by TBA for group III-V compounds, and group-IV combinations, respectively. The fitting procedure is succeeding for group III-V compounds. In the group-IV structures, some compounds ad-jointed tin such as monolayer of SnC, SnSi and SnGe have been confronted with the inefficiency of the TB method to the extent of the first nearest neighbor.
	
	\begin{table}[!h]
		\begin{center}
			\caption{\label{tab:2} Slater-Koster parameters of BX honeycomb lattice derived by TBA.}
			\begin{tabular}{cccccc}
				\hline \hline
				&BN&BP&BAs&BSb&BBi \\ \hline
				$\epsilon_{s_A}$   	& 0.359703	&-3.153820&-3.315744&-4.114327&-3.418540 \\ 
				$\epsilon_{P_{y_A}}$& 6.660174	&2.849500 &2.609751 &2.062226& -3.418540 \\ 
				$\epsilon_{P_{z_A}}$& 2.116069	&0.514927 &0.435839 &0.304630& 0.187126  \\ 
				$\epsilon_{P_{x_A}}$& 2.147729	&0.448072 &0.123560 &-0.120188&1.439313  \\ 		
				$\epsilon_{s_B}$    &-10.465849 &-6.591012&-7.395264&-6.215011&-8.434930 \\ 
				$\epsilon_{P_{y_B}}$& 0.506664	&1.166591 &1.155927 &1.406323& 0.609970  \\
				$\epsilon_{P_{z_B}}$& -2.329531	&-0.638155&-0.213239&-0.039482& -0.067821\\
				$\epsilon_{P_{x_B}}$& 2.150085 	&1.281628 &2.056722 &1.955744& 0.401624  \\			
				$V_{ss \sigma}$ 	& -4.364859	&-3.124032&-2.912592&-2.497178& -2.190991\\
				$V_{sp \sigma}$    	& 5.426549	&3.405441 &3.434569 &2.923374& 2.733426  \\
				$V_{pp \sigma}$     & 5.896137	&3.821502 &3.705766 &3.215200& 2.675127  \\
				$V_{pp \pi}$    	& -2.341442	&-1.469154&-1.548911&-1.340630& -1.256691\\
				\hline \hline
			\end{tabular}
		\end{center}
	\end{table}
	
	\begin{table}[!h]
		\begin{center}
			\caption{\label{tab:3} Slater-Koster parameters of the monolayer of IV-compounds derived by TBA.}
			\begin{tabular}{ccccccc}
				\hline \hline
				&Graphene&Silicene&Germanene&Stanene \\ \hline
				$\epsilon_{s_A}$   	&-3.204545&-4.342110&-5.7160&-4.819449 \\ 
				$\epsilon_{P_{y_A}}$&3.670414 &1.900000&3.398440&2.477134    \\ 
				$\epsilon_{P_{z_A}}$&0.0	  &0.0&0.0&0.0               \\ 
				$\epsilon_{P_{x_A}}$&2.690589 &1.451142&0.791040&1.443206   \\ 		
				$\epsilon_{s_B}$    &-&-&-&-  \\ 
				$\epsilon_{P_{y_B}}$&-&-&-&-   \\
				$\epsilon_{P_{z_B}}$&-&-&-&-  \\
				$\epsilon_{P_{x_B}}$&-&-&-&-   \\	
				$V_{ss \sigma}$ 	&-5.485425&-2.140001&-1.755720&-1.586277 \\
				$V_{sp \sigma}$    	&5.839044 &2.454669&2.418080&2.115327     \\
				$V_{pp \sigma}$     &6.479086 &2.630116&2.555960&2.231675    \\
				$V_{pp \pi}$    	&-2.749099&-1.113355&-0.86480&-0.927208   \\
				& &\\
				&SiC&SiGe&GeC&SnC \\  \hline 
				$\epsilon_{s_A}$   	&-7.836687&-5.313012&-8.573873&0.602780&   \\ 
				$\epsilon_{P_{y_A}}$ &-2.599611&2.837254&-2.431318&-2.052342&  \\ 
				$\epsilon_{P_{z_A}}$&1.114450&-0.029311&1.024802&-10.278306&  \\ 
				$\epsilon_{P_{x_A}}$&-3.620749&1.614479&-3.896555&-0.969448&   \\ 
				
				$\epsilon_{s_B}$    &-0.256636&-4.500626&-0.600946&-0.643479&  \\ 
				$\epsilon_{P_{y_B}}$&5.962847&0.932495&5.289562&19.998739&    \\
				$\epsilon_{P_{z_B}}$&-0.728725&0.0&-0.644296&1.346508&         \\
				$\epsilon_{P_{x_B}}$&3.524612&1.800000&3.511424&-0.308166&    \\
				
				$V_{ss \sigma}$ 	&-3.1421160&-2.161956&-3.068084&-1.038965&  \\
				$V_{sp \sigma}$    	&-3.591319&-2.651641&-3.578150&-1.754648&  \\
				$V_{pp \sigma}$     &-2.1297328&2.655390&-2.123841&3.018126&    \\
				$V_{pp \pi}$    	&1.870858&-1.077112&1.824391&-2.345869&   \\
				\hline \hline
			\end{tabular}
		\end{center}
	\end{table}
	
	As well as reported the Fermi velocity by DFT calculation at Table~\ref{tab:1}, we focus on the SiGe monolayer and by using this value will try to explore some physical aspects by tight-binding model. Obviously, other well-worked fitted results will be a topic for the continuation of this article.
	
	Based on the TBA mentioned at \ref{sec:AppA}, to understand the origin of the band-gaps in group III-V and group-IV materials, we can brief the TBA model considering the hopping matrix elements for interaction between the $p_z$ orbitals on nearest neighbor sites and neglecting any orbital overlaps. The Hamiltonian is
	
	\begin{equation}
		H=\left(\begin{array}{cc}
			\epsilon_{p_zA} & t \cdot g(k) \\
			t \cdot g(k)^{*} & \epsilon_{p_zB}
		\end{array}\right),
	\end{equation}
	
	where $p_zA$ and $p_zB$ are the onsite $p_z$ orbital energies of the A and B atoms, respectively. And $t$ is the hopping integral as $\frac{(\Delta^2 V_{pp\sigma}+V_{pp\pi})}{\sqrt{1+\Delta^2}}$, and $g(k)$ defined as
	
	\begin{equation}
		g(k)=e^{ik_z \Delta a}(e^{-ik_x a}+2cos(\frac{\sqrt{3}k_y a}{2})e^{i\frac{k_x a}{2}}),
	\end{equation}
	
	here $k_x$ and $k_y$ are the wavevector components. The secular equation, $det|H-E|=0$ , are solved for eigenvalues as 
	
	\begin{equation}
		E(k)=\frac{\epsilon_{p_zA}+\epsilon_{p_zB}}{2} \pm\left(\left(\frac{\epsilon_{p_zA}-\epsilon_{p_zA}}{2}\right)^{2}+(t \cdot g(k))^{2}\right)^{1 / 2}.
	\end{equation}
	
	At the Dirac point, $K$ hight symmetry point, neglecting buckled parameter compared to lattice constant, $g(k)^2=0$, so the energy band-gap is
	
	\begin{equation}
		E_{g}=\left|\epsilon_{p_zA}-\epsilon_{p_zB}\right|.
	\end{equation}
	
	This equation reveals that the energy band-gap depends on the difference of the energy between the $p_z$ orbitals of A and B. It is obvious that for materials with similar A and B atoms, such as graphene, we encounter zero band-gap.

	\subsection{\label{sec:BSb} Spin-orbit interaction and electric field in tight-binding model}

	The simple and effective $sp^3$ tight-binding model will be presented in Sec.~\ref{sec:AppA} in detail. The energy band structures have been gained, based on the Slater-Koster parameters shown at the Table~\ref{tab:2} and \ref{tab:3} for monolayers of group III-V and IV materials, respectively, we convinced that the results, means the band structure and band-gap, are in good agreement with the DFT results displaced in the Sec.~\ref{sec:Tpw}. In this section, we decide to peruse the topological aspects of these materials in the presence of tunning parameters like SOI and EF. So, with reference to this issue that topological feature of the band structure of insulators is considered around the Fermi level, we would like to examine the previous results, first in the presence of SOI, and second in the presence of EF.

	\begin{table}
		\begin{center}
			\caption{\label{tab:6} The band-gap (eV) values Vs. coefficients of initial SOI strength for a monolayer of BSb. For instance, $\times2$ means double in initial SOI value, and vice versa.}
			\begin{tabular}{ccccccccc}
				\hline \hline
				$\frac{SOI}{initial~SOI}$&&1.0&2.0&4.0&4.2&8.0 \\ \hline
				$\epsilon_g$& DFT &0.36&0.29&0.12&0.10&0.00           \\ 
				$(eV)$& TBA &0.33&0.30&0.18&0.17&0.09           \\ 
				$\lambda_{SO_{B}}$&   &0.05&$\times2$&$\times4$&$\times4.2$&$\times8$        \\
				$\lambda_{SO_{Sb}}$&  &0.50&$\times2$&$\times4$&$\times4.2$&$\times8$     \\
				\hline \hline
			\end{tabular}
		\end{center}
	\end{table}
	
\begin{figure}[!h]
		\centering
		\begin{tabular}[b]{c}
			\begin{minipage}[b]{0.55\textwidth}
				\includegraphics[width=\textwidth]{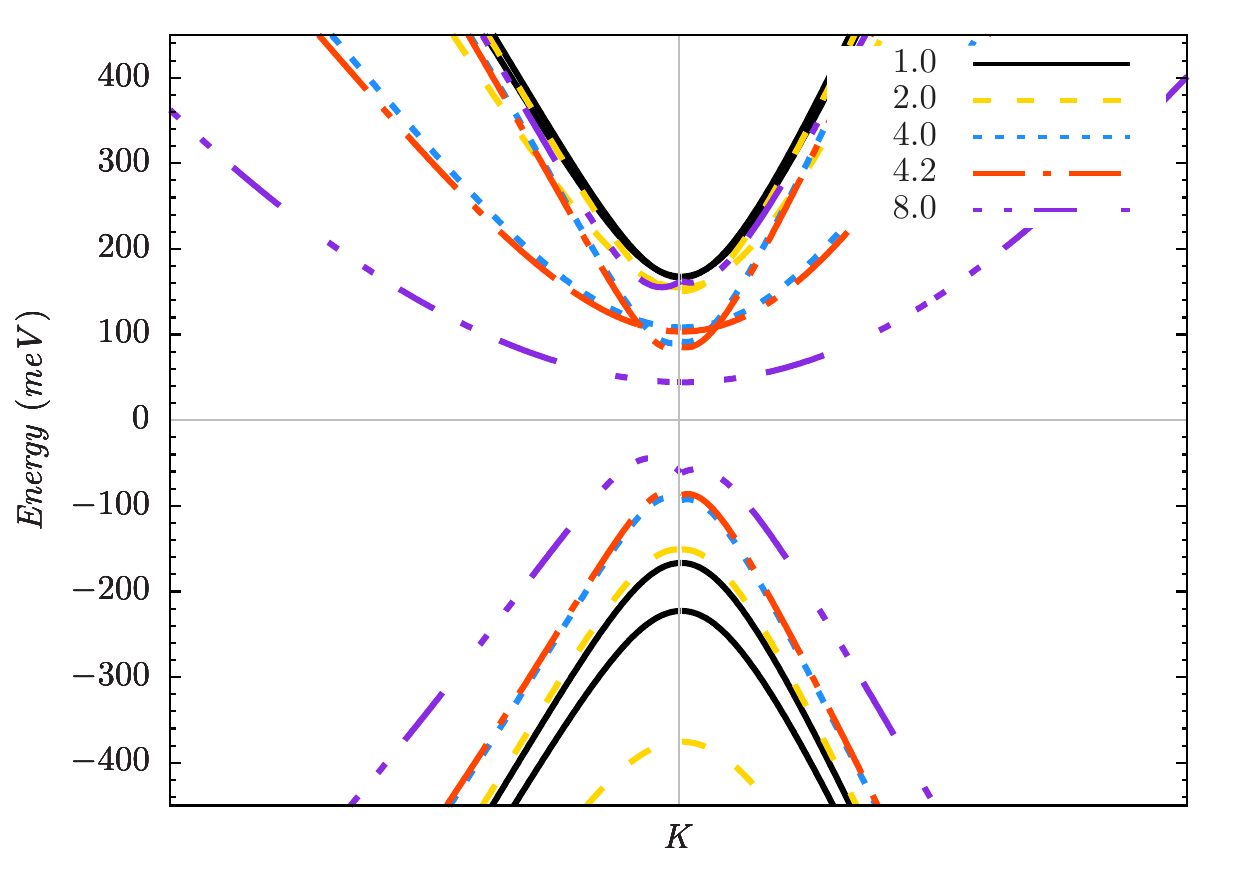}
			\end{minipage}\\
		\end{tabular}
		\vspace{-5.0mm}
		\caption{\label{fig:8}(color online).  TBA band structure for monolayer of BSb for coefficients of SOI strength: 1.0 (solid, black), 2.0 (yellow, dashed), 4.0 (blue, dotted) , 4.2 (red, dashed-dot-dashed) , 8.0 (magenta, dot-dot-dashed). }
	\end{figure}

	First, the band-gaps have been derived by TBA for a monolayer of BSb when the intrinsic SOI strength is changed, besides the similar DFT results illustrated in Table~\ref{tab:6}. It is obvious that the band-gap values descend as the SOI strength increases. Here, the main purpose is to obtain the same results for band structure with the TBA model. As Fig.~\ref{fig:8} shows, this is carried out by considering the Slater-Koster parameters, presented in Table~\ref{tab:2}, and changing the SOI coefficients of Boron and antimony, presented as $\lambda_{so}$ in Table~\ref{tab:6}. Therefore, the output for DFT and TBA has identical behavior that is concluded our model is good enough to describe the band structure and topological effects in the presence of SOI. This paves the way for further investigation by TBA to provide an effective Hamiltonian near the Fermi level around the $K$ high symmetry point.
	
	\begin{figure}[!h] 
		\centering
		\begin{tabular}[b]{c}
			\begin{minipage}[b]{0.55\textwidth}
				\includegraphics[width=\textwidth]{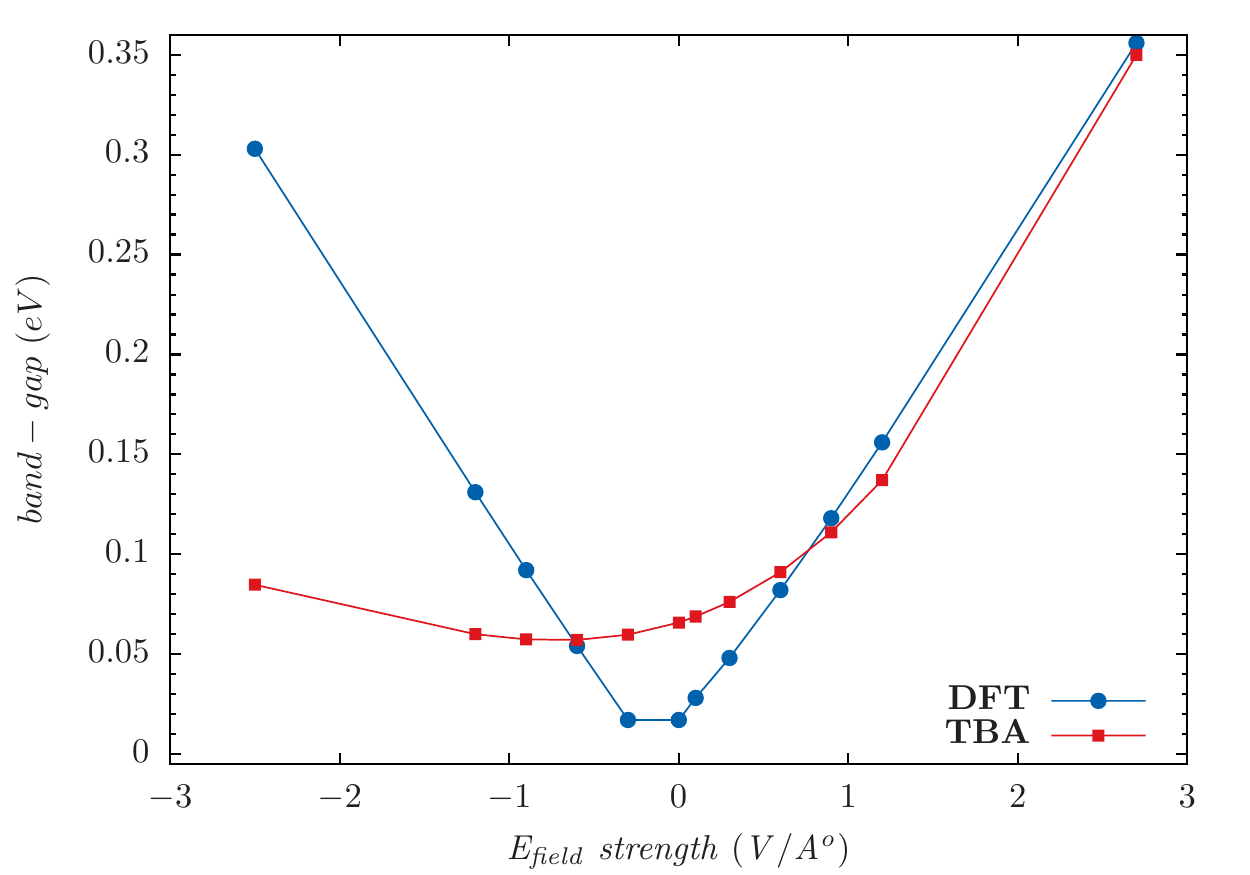}
			\end{minipage}\\
		\end{tabular}
		\vspace{-5.0mm}
		\caption{\label{fig:10}(color online). The monolayer of SiGe band-gap with TBA and DFT Vs. EF shows by square, red and diamond, black, respectively.}
	\end{figure}
	
\begin{figure}[!h]
		\centering
		\begin{tabular}[b]{c}
			\begin{minipage}[b]{0.55\textwidth}
				\includegraphics[width=\textwidth]{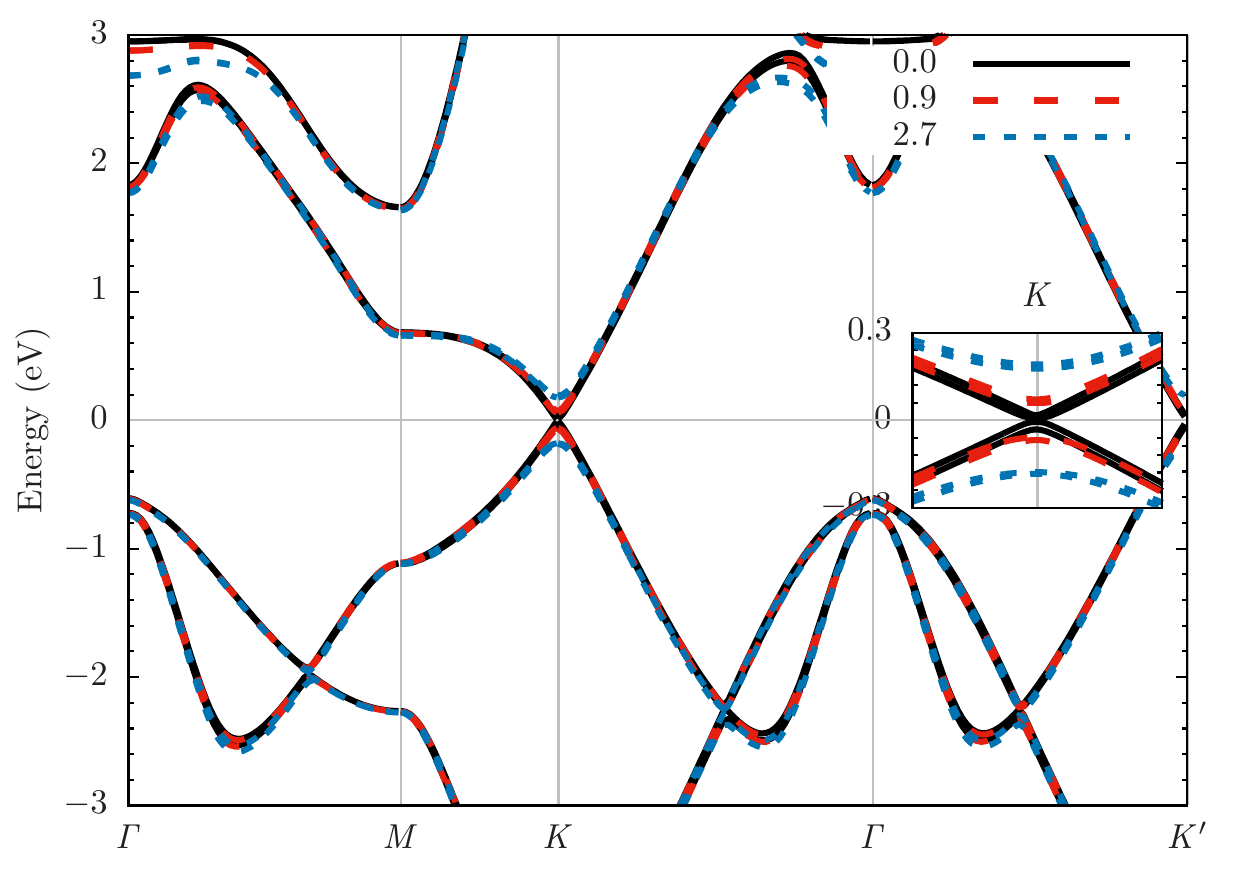}
			\end{minipage}\\
		\end{tabular}
		\vspace{-5.0mm}
		\caption{\label{fig:9}(color online).  TBA for the monolayer of SiGe for EF strength($\frac{V}{\AA{}}$) 0.0 (line, black), 0.9 (dashed, red), 2.7 (dotted, blue). }
	\end{figure}

	Second, the comparison between DFT and TBA about band-gap for SiGe monolayer in the presence of EF announced by Fig.~\ref{fig:9}, shows the band-gap increases by raising the EF strength. These values were presented in Table~\ref{tab:5}. At a border value of EF around $2.7~\frac{V}{\AA{}}$ is larger than $0.3~eV$, and the material enters the trivial phase. Our $sp^3$ tight-binding model Hamiltonian needs to expand overlap integrals in the presence of EF, so we need to use the Gaunt coefficients parameters \cite{Guseinov_1999,Guseinov_2003} to get the band structure. The first and most important terms after applying EF, are on-site contributions, named $\gamma_{sp}$, which is more discussed by C. R. Ast~\cite{Ast_2012_efield}. For each atoms, we have different $\gamma_{sp}$,
	
	\begin{equation}
		\begin{array}{ll}
			\gamma_{sp_{Si}}=-a_0 \frac{(2n^*_{Si}+1)n^*_{Si}}{2 \sqrt{3}(Z_{Si}-s_{Si})} ~=- 0.513103~\AA{},	\\
			\gamma_{sp_{Ge}}=-a_0 \frac{(2n^*_{Ge}+1)n^*_{Ge}}{2 \sqrt{3}(Z_{Ge}-s_{Ge})}=- 0.840035~\AA{},	\\
		\end{array}
	\end{equation}
	
	which in these relation $n^*$ is effective principle quantum number, 3 for Si and 3.7 for Ge. Z is supposed to be the actual charge in the nucleus, and s is a screening constant \cite{Slater_1930_STO}, 7.75 for Si and 26.35 for Ge. Moreover, other overlap integrals in the presence of EF, $(\gamma_{sp2}-\gamma_{sp1})$ and $\gamma_{pp1}$ calculated as 0.353372 \AA{} and 0.141243 \AA{}, respectively. The appropriate strengths of intrinsic SOI coefficients are $\lambda_{SO_{Si}}=0.014$ and $\lambda_{SO_{Ge}}=0.074$.
	
	Because of using the GGA functional in DFT calculation and underestimating band-gap size, an expectation of an exact matching between DFT and TBA result is far from reality, whether for a monolayer of BSb or SiGe, but the general behavior of both calculations is the same. As Fig.~\ref{fig:10} shows regardless of the magnitude of the band-gap, the behavior of its changes is consistent with the change in the intensity of the EF between DFT and TBA. So. it is obvious that the TBA model presented in the presence of EF has succeeded too.
	
	\section{\label{sec:Summary}Summary}
	
	In summary, based on the first-principles density functional theory, we have investigated atomic structures, electronic structures, and topological features of 2D binary materials, which composed of group III-V and IV-IV elements, with planar and low-buckled geometries like graphene ($sp^2$) and silicene ($sp^3$), respectively. Using the OpenMX package, which enabled us to manipulate spin-orbit interaction and apply a perpendicular electric field, we have explored the band-gap size, the existence of the Dirac point and the topological phase transition, which are affected by these parameters, in some compounds. In the case of the monolayer of BSb and SiGe, the band-gap closed, and these materials transitioned to topological and trivial insulator phases, respectively. As well as, the calculating $\mathbb{Z}_2$ invariant proved these phase transitions. The other main result, based on a $sp^3$ tight-binding model containing SOI and EF, is the set of Slater–Koster parameters, which have been obtained from a fit to DFT calculations. These accurately have reproduced the band structures and the topological patterns in the presence of spin-orbit interaction and external electric field. In this way, our results prepare a context for studying effective Hamiltonian and spintronic applications of 2D topological nanostructures and can be quite promising for nano-electronics.
	
	\section{\label{sec:Thanks}Acknowledgments}
	
	We would like to thank Dr. S. Mahdi Fazeli (Department of Physics, University of Qom, Qom, Iran) for his detailed scientific discussions and guidance in this article. We would like also to show our gratitude to Dr. Hyun-Jung Kim (Peter Grünberg Institut (PGI-1), Forschungszentrum Jülich, Germany) for helpful discussions. We also thank Prof. I.I. Guseinov (Çanakkale Onsekiz Mart Üniversitesi, Çanakkale, Turkey) for providing the initial program code of the overlap integral over Slater-type orbitals.
	
	\appendix
	\section{\label{sec:AppA}The tight-binding considerations}
	
	In this section, we will briefly introduce the tight-binding model has been used for 2D honeycomb lattice, planar like as graphene, and buckled structures like as silicene. As the Fig.~\ref{fig:1} shows, the atom of type A has three first nearest neighbor ($n.n.$) of the type B, and we neglect more further neighbors. The basis sets of two-atoms unit cell contain $A_0$ and $B_1$ are: $\vec{a}_1=b(\frac{3}{2},\frac{\sqrt{3}}{2}, \Delta \times \frac{a}{b}) $ and $\vec{a}_2=b(\frac{3}{2},-\frac{\sqrt{3}}{2}, \Delta \times \frac{a}{b})$, which $a$ is the lattice constant and set to unity, and $b=\frac{1}{\sqrt{3}} a$ is the distance of first $n.n.$. So if we consider the atom of type A at the origin ($A_0:(0,0,0)$), three first $n.n.$ (B atoms) located at $B_1:(-1,0,\Delta) b$ , $B_2:(\frac{1}{2},\frac{\sqrt{3}}{2},\Delta) b$ and $B_3:(\frac{1}{2},\frac{-\sqrt{3}}{2},\Delta) b$. Also, we consider the buckling parameter, $\Delta$, is lower than lattice constant, $\Delta << a=1$, and the structure is low-buckled. By using the Slater-Koster formalism \cite{Slater_1954_TB}; considering $s$ and $p$ orbitals of A and B atoms; and overlap between them; the $8 \times 8$ Hamiltonian in the presence of an EF perpendicular to the plane of lattice, will be as $H_{TB}$, where 
	
	\begin{equation}
		\begin{array}{lll}
			G_{0}=e^{-i k_{x}}e^{i k_{z}\Delta}, \\
			G_{1}=2 \cos \left(\frac{\sqrt{3} k_{y}}{2}\right) e^{i \frac{k_{x}}{2}}e^{i k_{z}\Delta}, \\
			G_{2}=2 i \sin \left(\frac{\sqrt{3} k_{y}}{2}\right) e^{i \frac{k_{x}}{2}}e^{i k_{z}\Delta} ,
		\end{array}
	\end{equation}
	
	\noindent are geometry functions, and 
	
	\begin{equation}
		\begin{array}{llll}
			A=\left(V_{p p \sigma}+3\Delta^2 V_{p p \pi}\right),\\ 
			B=\left(V_{p p \sigma}-V_{p p \pi}\right),\\ 
			C=\left(V_{p p \sigma}+(3+4\Delta^2)V_{p p \pi}\right),\\ 
			D=\left(3V_{p p \sigma}+(1+4\Delta^2)V_{p p \pi}\right),\\
			F=\left(\Delta^2 V_{p p \sigma}+V_{p p \pi}\right),\\
			R=\sqrt{1+\Delta^2},
		\end{array}
	\end{equation}
	
	\noindent A, B, C, D, F and R are the parameters.

	\begin{landscape}
		\tiny \begin{equation}
			H_{TB}=\left(\begin{array}{rccccccccc} 
				& S_{A} & P_{x_{A}} & P_{y_{A}} & P_{z_{A}} & S_{B} & P_{x_{B}} & P_{y_{B}} & P_{z_{B}} \\
				\\ \hline \\\\
				S_{A} \vline	& \epsilon_{s_{A}} & 0 & 0 & -a_0 E_{f} \frac{(2n_A+1)n_A}{2 \sqrt{3}(Z_A-s_A)} & V_{s s \sigma}\left(G_{0}+G_{1}\right) & \frac{V_{s p \sigma}\left(-G_{0}+\frac{1}{2} G_{1}\right)}{R} & \frac{V_{s p \sigma}\left(\frac{\sqrt{3}}{2} G_{2}\right)}{R} & \frac{\Delta V_{s p \sigma}\left(G_{0}+G_{1}\right)}{R} \\ \vline \\
				
				P_{x_{A}} \vline& 0 & \epsilon_{p_{x_{A}}} & 0 & 0 & \frac{V_{s p \sigma}\left(G_{0}-\frac{1}{2} G_{1}\right)}{R} & \frac{A \left(G_{0}\right)+\frac{C}{4} G_{1}}{R^2} & \frac{\frac{\sqrt{3}}{4}B G_{2}}{R^2} & \frac{\Delta B\left(-G_{0}+\frac{1}{2} G_{1}\right)}{R^2} \\ \vline \\
				
				P_{y_{A}} \vline& 0 & 0 & \epsilon_{p_{y_{A}}} & 0 & -\frac{V_{s p \sigma}\left(\frac{\sqrt{3}}{2} G_{2}\right)}{R} & \frac{\frac{\sqrt{3}}{4}B G_{2}}{R^2} & V_{p p \pi}\left(G_{0}\right)+\frac{D  G_{1}}{4R^2} & \frac{\frac{\sqrt{3}}{2}\Delta B G_{2}}{R^2} \\ \vline \\
				
				P_{z_{A}} \vline& -a_0 E_{f} \frac{(2n_A+1)n_A}{2 \sqrt{3}(Z_A-s_A)} & 0 & 0 & \epsilon_{p_{z_{A}}} & -\frac{\Delta V_{s p \sigma}\left(G_{0}+G_{1}\right)}{R} & \frac{\Delta B\left(-G_{0}+\frac{1}{2} G_{1}\right)}{R^2} & \frac{\frac{\sqrt{3}}{2}\Delta D G_{2}}{R^2} & \frac{F(G_{0}+G_{1})}{R^2} \\ \vline \\

				S_{B} 	 \vline & V_{s s \sigma}\left(G^\dagger_{0}+G^\dagger_{1}\right) & \frac{V_{s p \sigma}\left(G^\dagger_{0}-\frac{1}{2} G^\dagger_{1}\right)}{R} & -\frac{V_{s p \sigma}\left(\frac{\sqrt{3}}{2} G^\dagger_{2}\right)}{R} & -\frac{\Delta V_{s p \sigma}\left(G^\dagger_{0}+G^\dagger_{1}\right)}{R} & \epsilon_{s_{B}} & 0 & 0 &  -a_0 E_{f} \frac{(2n_B+1)n_B}{2 \sqrt{3}(Z_B-s_B)} \\ \vline \\
				
				P_{x_{B}} \vline& \frac{V_{s p \sigma}\left(-G^\dagger_{0}+\frac{1}{2} G^\dagger_{1}\right)}{R} & \frac{A \left(G^\dagger_{0}\right)+\frac{C}{4} G^\dagger_{1}}{R^2} & \frac{\frac{\sqrt{3}}{4}B G^\dagger_{2}}{R^2} &\frac{\Delta B\left(-G^\dagger_{0}+\frac{1}{2} G^\dagger_{1}\right)}{R^2} &0&\epsilon_{p_{x_{B}}} & 0 & 0\\ \vline \\
				
				P_{y_{B}} \vline&\frac{V_{s p \sigma}\left(\frac{\sqrt{3}}{2} G^\dagger_{2}\right)}{R} & \frac{\frac{\sqrt{3}}{4}B G^\dagger_{2}}{R^2} & V_{p p \pi}\left(G^\dagger_{0}\right)+\frac{D  G^\dagger_{1}}{4R^2} &\frac{\frac{\sqrt{3}}{2}\Delta D G^\dagger_{2}}{R^2} & 0 & 0 &\epsilon_{p_{y_{B}}} & 0 \\ \vline \\
				
				P_{z_{B}} \vline& \frac{\Delta V_{s p \sigma}\left(G^\dagger_{0}+G^\dagger_{1}\right)}{R} & \frac{\Delta B\left(-G_{0}+\frac{1}{2} G^\dagger_{1}\right)}{R^2} &  \frac{\frac{\sqrt{3}}{2}\Delta B G^\dagger_{2}}{R^2} & \frac{F(G^\dagger_{0}+G^\dagger_{1})}{R^2} & -a_0 E_{f} \frac{(2n_B+1)n_B}{2 \sqrt{3}(Z_B-s_B)} & 0 & 0 & \epsilon_{p_{z_{B}}}
			\end{array}\right).
		\end{equation} \\ \\
		
		\normalsize
		
	\end{landscape}

	The perpendicular EF effect consider with $\lambda_{E_{f}}$ as strength; $n_A(n_B)$ and $Z_A(Z_B)$ are effective quantum number and atomic number of atom A(B), respectively; and $a_0$ is Bohr radius. The on-site terms calculated as $\gamma_{sp}=<s|E_fz|p_z>$ by considering $z$ as $rcos(\theta)$. Also, for orbital functions, we used the Slater spherical form of them; and $n$ as effective quantum number; and $s$ as screening constant~\cite{Slater_1930_STO}. Then other off-site terms related to overlap between neighbor atoms consider as $H_{z,off-site}$, with $l,m,n$ as directional cosines depend on the connecting vector.
	
	\begin{equation}
		\begin{array}{llll}
			\gamma_{sp1}=<s|z|p_z,b \hat{x}>,\\ 
			\gamma_{sp2}=<s,-\frac{\Delta}{2} \hat{z}|z|p_z,\frac{\Delta}{2} \hat{z}>,\\  
			\gamma_{pp1}=<p_x|z|p_z,b \hat{x}>, 
		\end{array}
	\end{equation}\\
	
	\noindent and 
	
	\begin{equation}
		\begin{tabular}{ccc}
			$H_{z,off-site}=$ & $ E_f \left( \begin{tabular}{c}0\\$\frac{(l_1n_1G_0+l_2n_2G_2)\left(\gamma_{sp2}-\gamma_{sp1}\right)}{R^2}$\\$\frac{(m_1n_1G_0+m_2n_2G_2)\left(\gamma_{sp2}-\gamma_{sp1}\right)}{R^2}$\\$\frac{(\left(1-n_1^{2}\right) \gamma_{sp1}+n_1^{2} \gamma_{sp2})(G_0+G_1)}{R^2}$\\\end{tabular} \right.$ & $\begin{tabular}{c}$\frac{(l_1n_1G_0+l_2n_2G_2)\left(\gamma_{sp2}-\gamma_{sp1}\right)}{R^2}$\\0\\0\\$-\frac{(l_1G_0+l_2G_1) \gamma_{pp1}}{R^2}$\\\end{tabular}$ \\ \\
			& $\begin{tabular}{c}$\frac{(m_1n_1G_0+m_2n_2G_2)\left(\gamma_{sp2}-\gamma_{sp1}\right)}{R^2}$\\0\\0\\$-\frac{(m_1G_0+m_2G_2) \gamma_{pp1}}{R^2}$\\\end{tabular}$ & $\left. \begin{tabular}{c}$\frac{(\left(1-n_1^{2}\right) \gamma_{sp1}+n_1^{2} \gamma_{sp2})(G_0+G_1)}{R^2}$\\$\frac{(l_1G_0+l_2G_1) \gamma_{pp1}}{R^2}$\\$\frac{(m_1G_0+m_2G_2) \gamma_{pp1}}{R^2}$\\0\\\end{tabular} \right)$,\\
		\end{tabular}
	\end{equation}
	
	\noindent more discussion prepared by C. R. Ast~\cite{Ast_2012_efield}.
	
	The SOI in the tight-binding method is traditionally taken into account only for $p$ orbitals on interatomic matrix elements of the Hamiltonian \cite{Chadi_1975_TB}. We use it in the form of the below matrix:
	
	\begin{equation}
		H_{SOI}=\lambda_{so}\left(\begin{array}{rcccccc} 
			& P_{x_{A}}^\uparrow & P_{x_{A}}^\downarrow & P_{y_{A}}^\uparrow & P_{y_{A}}^\downarrow & P_{z_{A}}^\uparrow & P_{z_{A}}^\downarrow \\ \hline 
			P_{x_{A}}^\uparrow \vline	&0&0&-$i$&0&0&1  \\
			P_{x_{A}}^\downarrow \vline &0&0&0&$i$&-1&0   \\ 
			P_{y_{A}}^\uparrow \vline   &$i$&0&0&0&0&-$i$     \\ 
			P_{y_{A}}^\downarrow \vline &0&-$i$&0&0&-$i$&0   \\ 
			P_{z_{A}}^\uparrow 	 \vline &0&-1&0&$i$&0&0  \\
			P_{z_{A}}^\downarrow \vline &1&0&$i$&0&0&0  \\
		\end{array}\right).
	\end{equation}  \\
	
	Finally the overall Hamiltonian matrix is $H=H_{TB}~(include~H_{z,off-site})+H_{SOI}$. It is obvious that when the SOI consider, the Hamiltonian size changes to double ($16 \times 16$).
	
	\vspace{15pt}
	\bibliography{abaradaranbib}
	
\end{document}